\documentclass[preprint,prc,aps,showpacs,amsmath,nofootinbib]{revtex4}

\begin{document}
\title{Spin structure of spin-1/2 baryon and spinless meson 
production amplitudes in photo and hadronic reactions.}
\author{K. Nakayama, W. G. Love}
\affiliation{Department of Physics and Astronomy, University of Georgia, 
Athens, GA 30602, USA }

\begin{abstract}
The most general spin structures of the spin-1/2 baryon and spinless meson production 
operator for both photon and nucleon induced reactions are derived from the partial-wave 
expansions of these reaction amplitudes. The present method provides the coefficients 
multiplying each spin operator in terms of the partial-wave matrix elements. The 
result should be useful in studies of these reactions based on partial-wave analyses, 
especially, when spin observables are considered. 
\end{abstract}

\vskip 0.5cm
\maketitle
%
\newpage
In the present work, we derive the most general spin structure of the reaction 
amplitude for both positive and negative parity spin-1/2 baryon production in 
nucleon-nucleon ($NN$) collisions and also in photon induced processes on nucleons. 
Knowing the spin structure of the transition operator is of particular importance
in analyses of spin observables. The method used here to extract the spin 
structure is a generalization of the partial-wave expansion of the $NN$ amplitude 
following Ref.\cite{NL}. It is quite general and, in principle can be applied to 
any reaction process in a systematic way. Usually, the structure of a transition 
operator is derived based solely on symmetry principles. The usefulness of the 
present method is that it also provides explicit formulae for the coefficients 
multiplying each spin structure in terms of the partial-wave matrix elements and 
this should be particularly useful in model-independent analyses based on the 
partial-wave expansion of the reaction amplitude. 

As an example of application of the present method, we investigate the possibility
of determining the parity  of a narrow resonance in the 
photon as well as in the nucleon induced reactions. The search for new resonances, 
especially the so-called missing resonances and exotic resonances, are receiving 
increased attention \cite{exotics}. Apart from establishing their existence, the 
determination of their basic properties are of extreme 
importance. Among these properties, the parity is of particular interest in 
connection with the substructure of these resonances \cite{Jennings}. 
However, it is often the case that no theoretical predictions can provide a 
conclusive result for the parity and other basic properties. A recent example of this
situation is provided by the pentaquark $\Theta^+$. The existence of this exotic 
baryon and the determination of its spin and parity quantum numbers have been under 
an intensive investigation both experimentally and theoretically over the past 
couple of years \cite{theta}
\footnote{The existence of the $\Theta^+$ is still controversial. While this resonance 
has been seen in a number of experiments, it has not been observed in a similar number 
of experiments, including a very recent experiment \cite{JLAB} with much better 
statistics.}. 
It is, therefore, very important to find a way of determining these properties 
in a model independent way.   

In the past few years a considerable amount of data for meson production in $NN$ 
collisions have been obtained (see Ref.~\cite{Moskal} for a review). In particular 
there are data, not only for cross sections but, in the case of pion production in 
$NN\to NN\pi$, a large set of spin observable data \cite{Meyer}. In addition, the 
data bases for the production of other mesons, such as $\eta$ and $\eta^\prime$, 
are growing rapidly \cite{Moskal,eta-etap}. Also, reactions involving the production 
of particles containing strange quarks such as the $NN\to YNK$ reaction where $Y$ 
stands for a hyperon, are receiving increased attention \cite{Moskal,YNK}. With 
these (high-precision) data becoming available, there is 
a demand for more thorough and detailed theoretical analyses of these reactions. 
Therefore, in the present work, we also derive the most general spin-isospin structure 
of the $NN \to NB'M$ transition operator, where $M$ stands for a spinless (scalar or 
pseudoscalar) meson and, $B'$, for a spin-1/2 baryon with positive-parity. The spin 
structure of the $NN \to NN\pi$ transition operator has been derived in Ref.\cite{Deepak} 
and applied to neutral pion production. Here, we also consider charged meson production.

The present paper is organized as follows: In section I, the general spin structure 
of the transition operator is derived for the reaction $\gamma + N \to M + B$, where 
$B$ stands for a spin-1/2 and either a positive or negative parity baryon. In section 
II the results derived in section I are 
illustrated by applying them to the near-threshold kinematic regime.
Sections III and IV are devoted to the derivation of the most general spin 
structure of the reaction $N + N \to B' + B$, where $B'$ stands for a spin-1/2 and 
positive-parity baryon, and to its application near-threshold, respectively. The reaction 
$N + N \to M + B' + N$ is considered in sections V and VI. A summary is given in section 
VII. Appendices A-D contain some details of the derivation of the spin structure of the 
transition operators.

\section{The reaction $\gamma + N \to M + B$.}

We start by making a partial-wave expansion of the $\gamma + N \to M + B$ reaction 
amplitude. Here $M$ stands for a pseudoscalar meson and $B$, a spin-1/2 baryon. 
We, then, have
\begin{eqnarray}
<\frac{1}{2}m'|\hat M(\vec q, \vec k)|SM_S>  = 
& \sum & i^{L-L'} (SM_SLM_L|JM_J)(\frac{1}{2}m'L'M_{L'}|JM_J)
M^{JS}_{L'L}(q,k) \nonumber \\
& \times & Y_{L'M_{L'}}(\hat q) Y^*_{LM_L}(\hat k) \ ,  
\label{PWE1}
\end{eqnarray}
where $S, L, J$ stand for the total spin, total orbital angular momentum and the 
total angular momentum, respectively, of the initial $\gamma N$ state. $M_S$, $M_L$
and $M_J$ denote the corresponding projection quantum numbers. The primed quantities 
stand for the corresponding quantum numbers of the final $MB$ state. 
The summation runs over all quantum numbers not specified in the left-hand side 
(l.h.s.) of Eq.(\ref{PWE1}). $\vec k$ and $\vec q$ denote the relative momenta of 
the two particles in the initial and final states, respectively. The partial-wave 
expansion given above is, of course, related to the more commonly used electric and 
magnetic multipole expansion. For the present analysis, however, it is convenient to 
use the above expansion.

Eq.(\ref{PWE1}) can be inverted to solve for the partial-wave matrix element
$M^{SJ}_{L'L}(q, k)$. We have
\begin{eqnarray}
M^{SJ}_{L'L}(q, k) & = &
\sum i^{L'-L} (\frac{1}{2}m'L'M_{L'}|JM_{J})(SM_SL0|JM_J)\nonumber \\
& \times & \frac{8\pi^2}{2J+1} \sqrt{\frac{2L+1}{4\pi}}
\int_{-1}^{+1} d(\cos(\theta)) 
Y_{L'M_{L'}}^*(\theta, 0)
<\frac{1}{2}m'|\hat M(\vec q, \vec k)|SM_S> \ ,
\label{pwme1}
\end{eqnarray}
where, without loss of generality,  $\vec k$ is chosen along the $z$-axis and  
$\vec q$ in the $xz$-plane; $\cos(\theta) \equiv \hat q \cdot \hat k$. The summation 
is over all quantum numbers not specified in the l.h.s. of the equation.

The most general spin structure of the transition operator can be extracted from 
Eq.(\ref{PWE1}) as
\begin{equation}
\hat M(\vec q, \vec k) = \sum_{SM_{S}m'} 
|\frac{1}{2}m'> <\frac{1}{2}m'|\hat M(\vec q, \vec k)|SM_S> <SM_S| \ .
\label{SSTRUC01}
\end{equation}
Inserting Eq.(\ref{PWE1}) into Eq.(\ref{SSTRUC01}) and re-coupling gives
\begin{equation}
\hat M(\vec q, \vec k) = \sum i^{L-L'} (-)^{-J-\frac{1}{2}} [J]^2  M^{JS}_{L'L}(q,k)
\sum_\alpha
  \begin{Bmatrix}
  S & L & J \\   L' & \frac{1}{2} & \alpha
  \end{Bmatrix}
[B_S \otimes A_{\frac{1}{2}}]^\alpha \cdot 
[Y_L(\hat k) \otimes Y_{L'}(\hat q)]^\alpha \ ,
\label{SSTRUC11}
\end{equation}
where we have used the notations $B_{SM_{S}}\equiv (-)^{S-M_S}<S-M_S|$, 
$A_{\frac{1}{2}m'}\equiv |\frac{1}{2}m'>$, and $[J]\equiv \sqrt{2J+1}$. 
The outer summation is over the quantum numbers $S$, $L$, $L'$, and $J$. In the above 
equation $S$ is either 1/2 or 3/2 so that $\alpha$ takes the values 0, 1, and 2, 
and denotes the rank of the corresponding tensor. In the above equation it should be
understood that the matrix elements of the meson creation and photon annihilation 
operators have already been taken.

We now expand $[B_S \otimes A_{\frac{1}{2}}]^\alpha$, for each tensor of rank $\alpha$, 
in terms of the complete set of available spin operators in the problem, i.e., the 
photon polarization vector $\vec\epsilon$ and the Pauli spin matrix $\vec \sigma$ 
together with the identity matrix. The result is
\begin{eqnarray}      
[B_S \otimes A_{\frac{1}{2}}]^0 & = & \frac{1}{\sqrt{6}}\ \vec\sigma \cdot 
\vec\epsilon \ ,\nonumber \\
\small[B_S \otimes A_{\frac{1}{2}}\small]^1 & = & - \frac{1+\sqrt{2}}{\sqrt{6}} \  
\left\{ \vec\epsilon + \frac{i}{\sqrt{2}} \left( \frac{1-\sqrt{2}}{1+\sqrt{2}}  
\right) \ (\vec\sigma \times \vec\epsilon) \right\}  \ , \nonumber \\
\small[B_S \otimes A_{\frac{1}{2}}\small]^2 & = & \frac{1}{\sqrt{2}} 
[\vec\sigma \otimes \vec\epsilon]^2 \ ,
\label{SOPER1}
\end{eqnarray}
where the numerical factors are uniquely determined such that the spin matrix 
elements of the right-hand side (r.h.s.) in the above equations equal the 
corresponding matrix elements of the l.h.s.

What we have done so far applies to either a negative or positive parity baryon 
$B$. Total parity conservation demands that $(-)^{L+L'}=+1$ and 
$(-)^{L+L'}=-1$ in the case of a positive and negative parity $B$, respectively. 
This leads to distinct spin structures of the transition operator for positive 
and negative parity as we shall show below. Hereafter, the superscript $\pm$ on 
any quantity stands for the positive $(+)$ or negative $(-)$ parity of $B$.

\subsection{Positive parity case}
For a positive parity baryon $B$, choosing the quantization axis $\hat z$ along 
$\hat k$, the quantity $[Y_L(\hat k) \otimes Y_{L'}(\hat q)]^\alpha$ can be 
expressed  without loss of generality as
\begin{eqnarray}
[Y_L(\hat k) \otimes Y_{L'}(\hat q)]^0 & = & (-)^L \frac{[L]}{4\pi} P_L(\hat q 
\cdot \hat k) \delta_{L,L'} \ , \nonumber \\
\small[Y_L(\hat k) \otimes Y_{L'}(\hat q)\small]^1 & = & i (-)^L \frac{[L]}{4\pi}
\sqrt{\frac{3}{L(L+1)}} P^1_L(\hat q \cdot \hat k) \delta_{L,L'} \hat n_2 \ , 
\nonumber \\
\small[Y_L(\hat k) \otimes Y_{L'}(\hat q)\small]^2 & = & 
a_{L'L} [\hat q \otimes \hat q]^2 + b_{L'L} [\hat k \otimes \hat k]^2 
+ c_{L'L} [\hat k \otimes \hat q]^2  \ .
\label{aux11}  
\end{eqnarray}
The structure in the above equation is dictated by total parity conservation. 
$P_L (P_L^1)$ is the ordinary (associated) Legendre function. 
$\hat n_2 \equiv (\hat k \times \hat q)/|\hat k \times \hat q|$. The coefficients 
$a_{L'L}$, $b_{L'L}$ and $c_{L'L}$ are derived explicitly in Appendix A.

Inserting Eqs.(\ref{SOPER1},\ref{aux11}) into Eq.(\ref{SSTRUC11}) we have
\begin{eqnarray}
\hat M^+(\vec q, \vec k) & = &
{\cal F}_1 \vec\sigma\cdot\vec\epsilon + i{\cal F}_2 \vec\epsilon \cdot\hat n_2 
+ {\cal F}_3 (\vec\sigma\times\vec\epsilon)\cdot\hat n_2 \nonumber \\
& + &
  {\cal F}_4 [\vec\sigma\otimes\vec\epsilon]^2\cdot[\hat q \otimes \hat q]^2 
+ {\cal F}_5 [\vec\sigma\otimes\vec\epsilon]^2\cdot[\hat k \otimes \hat k]^2 
+ {\cal F}_6 [\vec\sigma\otimes\vec\epsilon]^2\cdot[\hat k \otimes \hat q]^2 
\ , 
\label{SSTRUC12}
\end{eqnarray}
where
\begin{eqnarray}
{\cal F}_1 & = & \frac{1}{8\pi\sqrt{3}} \sum [J]^2  M^{J\frac{1}{2}}_{LL}(q,k) 
P_L(\hat k \cdot \hat q) \ , \nonumber \\
{\cal F}_2 & = & -\frac{1}{4\pi} \left(\frac{1+\sqrt{2}}{\sqrt{2}}\right) 
\sum (-)^{-J-\frac{1}{2}+L} [J]^2 \frac{[L]}{\sqrt{L(L+1)}}
  \begin{Bmatrix}
  S & L & J \\   L & \frac{1}{2} & 1
  \end{Bmatrix}
M^{JS}_{LL}(q,k) P^1_L(\hat k \cdot \hat q) \ , \nonumber \\ 
{\cal F}_3 & = & \frac{1}{\sqrt{2}} \left(\frac{\sqrt{2}-1}{\sqrt{2}+1}\right) 
{\cal F}_2 \ ,  \nonumber \\ 
{\cal F}_4 & = & \sqrt{\frac{1}{2}} \sum i^{L-L'}(-)^{-J-\frac{1}{2}}[J]^2 
  \begin{Bmatrix}
  \frac{3}{2} & L & J \\   L' & \frac{1}{2} & 2
  \end{Bmatrix}
M^{J\frac{3}{2}}_{L'L}(q,k) a_{L'L} \ , \nonumber \\
{\cal F}_5 & = & \sqrt{\frac{1}{2}} \sum i^{L-L'}(-)^{-J-\frac{1}{2}}[J]^2 
  \begin{Bmatrix}
  \frac{3}{2} & L & J \\   L' & \frac{1}{2} & 2
  \end{Bmatrix}
M^{J\frac{3}{2}}_{L'L}(q,k) b_{L'L} \ , \nonumber \\
{\cal F}_6 & = & \sqrt{\frac{1}{2}} \sum i^{L-L'}(-)^{-J-\frac{1}{2}}[J]^2 
  \begin{Bmatrix}
  \frac{3}{2} & L & J \\   L' & \frac{1}{2} & 2
  \end{Bmatrix}
M^{J\frac{3}{2}}_{L'L}(q,k) c_{L'L} \ . 
\label{SSTRUC12a}
\end{eqnarray}
The summations are over all quantum numbers appearing explicitly in the r.h.s. of
the equalities.

The above result is the most general form of the spin structure of the transition
operator consistent with symmetry principles. The coefficients ${\cal F}_j$ are 
functions of the energy of the system and scattering angle $\theta$ of the meson $M$ 
relative to the photon direction. We note that the transition operator in 
Eq.(\ref{SSTRUC12}) has six independent spin operators and is valid for both real
and virtual photons. Using the identity 
\begin{equation}
3 [\ \vec\sigma \otimes \vec\epsilon\ ]^2 \cdot [\ \hat a \otimes 
\hat b\ ]^2 = \frac{3}{2} [\ \vec\sigma \cdot \hat a \vec\epsilon \cdot \hat b 
+ \vec\sigma \cdot \hat b \vec\epsilon \cdot \hat a \ ] 
- (\hat a \cdot \hat b) \vec\sigma \cdot \vec\epsilon \ ,
\label{tensor}
\end{equation}
where $\hat a$ and $\hat b$ stand for arbitrary unit vectors, Eq.(\ref{SSTRUC12})
can be rewritten as
\begin{equation}
\hat M^+(\vec q, \vec k) = 
F_1 \vec\sigma\cdot\vec\epsilon + iF_2 \vec\epsilon \cdot\hat n_2 
+ F_3 \vec\sigma\cdot\hat k\vec\epsilon\cdot\hat q 
+ F_4 \vec\sigma\cdot\hat q\vec\epsilon\cdot\hat q 
+ F_5 \vec\sigma\cdot\hat k \vec\epsilon\cdot\hat k
+ F_6 \vec\sigma\cdot\hat q \vec\epsilon\cdot\hat k \ ,
\label{SSTRUC12_1P}
\end{equation}
where
\begin{eqnarray}
F_1 & = & {\cal F}_1 - \frac{1}{3} \Big[ {\cal F}_4 + {\cal F}_5  
      + (\hat q \cdot \hat k){\cal F}_6 \Big] \ , \ \ 
F_2 =  {\cal F}_2 \ , \ \
F_3 =  \frac{1}{|\hat k\times\hat q|}{\cal F}_3 + \frac{1}{2}{\cal F}_6  \ , \nonumber \\
F_4 & = & {\cal F}_4 \ , \ \ 
F_5 = {\cal F}_5 \ , \ \
F_6 =  - \frac{1}{|\hat k\times\hat q|}{\cal F}_3 + \frac{1}{2}{\cal F}_6 \ .
\label{SSTRUC12_1Pa}
\end{eqnarray}
It should be noted that Eq.(\ref{SSTRUC12_1P}) is equivalent to that of Ref.\cite{AG}. 
In fact, apart from an irrelevant overall factor of $i$, the coefficients 
$f_j \equiv {\cal F}^{V (\pm, 0)}_j, (j=1,...,6)$ in Ref.\cite{AG} are related to 
those in Eq.(\ref{SSTRUC12_1P}) by
\begin{eqnarray}
F_1 & = & f_1 - (\hat q\cdot\hat k)f_2 \ , \ \
F_2 = |\hat k\times\hat q|f_2 \ , \ \
F_3 = (f_2 + f_3) \ , \nonumber \\
F_4 & = & f_4 \ , \ \
F_5 = - f_1 - (\hat q \cdot \hat k)f_3 
                 - \frac{k^2}{k_0} f_5  \ , \ \
F_6 = - (\hat q\cdot\hat k) f_4 - \frac{k^2}{k_0} f_6 \ .
\label{AG-NL}
\end{eqnarray}

In the case of a real photon (photoproduction amplitude), the number of 
independent spin operators in Eq.\ref{SSTRUC12_1P} reduces to four due to the 
transversality condition, i. e., the terms $F_5$ and $F_6$ are absent. 
Eq.\ref{SSTRUC12_1P} then becomes equivalent to that of Ref.\cite{CGLN}. 
The coefficients $f_j, (j=1,...,4)$ in Ref.\cite{CGLN} are related to the 
corresponding coefficients $F_j$ in Eq.(\ref{SSTRUC12_1P}) as given by 
Eq.(\ref{AG-NL}) with $k^2=\vec\epsilon\cdot\hat k=0$.

One difference between our work [Eq.(\ref{SSTRUC12_1P})] and Refs.\cite{AG,CGLN} is 
that in the present work these coefficients are related explicitly 
[Eq.(\ref{SSTRUC12a})] to the partial-wave matrix elements introduced in 
Eq.(\ref{PWE1}).

\subsection{Negative parity case}
For a negative parity baryon $B$, choosing the quantization axis $\hat z$ along 
$\hat k$, $[Y_L(\hat k) \otimes Y_{L'}(\hat q)]^\alpha$ can be expressed as
\begin{eqnarray}
[Y_L(\hat k) \otimes Y_{L'}(\hat q)]^0 & = & 0 \ , \nonumber \\
\small[Y_L(\hat k) \otimes Y_{L'}(\hat q)]^1 & = & \frac{[LL']}{4\pi} 
\left[ \sqrt{\frac{2}{L'(L'+1)}}(L0L'1|11)P^1_{L'}(\hat k \cdot \hat q) \hat n_1 +
(L0L'0|10) P_{L'}(\hat k \cdot \hat q) \hat k \right] \ , \nonumber \\
\small[Y_L(\hat k) \otimes Y_{L'}(\hat q)]^2 & = & 
 a'_{L'L} [\hat k \otimes \hat n_2]^2 + b'_{L'L} [\hat q \otimes \hat n_2]^2 \ ,
\label{aux11n}  
\end{eqnarray}
where $\hat n_1 \equiv [(\hat k \times \hat q) \times \hat k]/|\hat k 
\times \hat q|$. The coefficients $a'_{L'L}$ and $b'_{L'L}$ are 
calculated explicitly in Appendix A.

Inserting Eqs.(\ref{SOPER1},\ref{aux11n}) into Eq.(\ref{SSTRUC11}) we have
\begin{eqnarray}
\hat M^-(\vec q, \vec k) & = &
i {\cal G}_1 \vec\epsilon\cdot\hat n_1 + {\cal G}_2 \vec\sigma \cdot 
(\vec\epsilon \times \hat n_1) 
+ {\cal G}_3 \vec\sigma \cdot (\vec\epsilon \times \hat k) \nonumber \\
& + &
   {\cal G}_4 [\vec\sigma\otimes\vec\epsilon]^2\cdot[\hat k \otimes \hat n_2]^2 
+  {\cal G}_5 [\vec\sigma\otimes\vec\epsilon]^2\cdot[\hat q \otimes \hat n_2]^2 
+ i {\cal G}_6 \vec\epsilon\cdot\hat k \ , 
\label{SSTRUC12n}
\end{eqnarray}
where
\begin{eqnarray}
{\cal G}_1 & = & \frac{1}{4\pi} \left( \frac{1+\sqrt{2}}{\sqrt{3}} \right) 
\sum i^{L-L'+1}(-)^{-J-\frac{1}{2}} [J]^2 \frac{[LL']}{\sqrt{L'(L'+1)}}(L0L'1|11) 
  \begin{Bmatrix}
  S & L & J \\   L' & \frac{1}{2} & 1
  \end{Bmatrix}
\nonumber \\ 
& \times &
M^{JS}_{L'L}(q,k) P^1_{L'}(\hat k \cdot \hat q) \ , \nonumber \\
{\cal G}_2 & = & \frac{1}{\sqrt{2}} \left(\frac{\sqrt{2}-1}{\sqrt{2}+1}\right) 
{\cal G}_1 \ , \nonumber \\ 
{\cal G}_3 & = & \frac{1}{\sqrt{2}} \left(\frac{\sqrt{2}-1}{\sqrt{2}+1}\right) 
{\cal G}_6 \ , \nonumber \\ 
{\cal G}_4 & = & \sqrt{\frac{1}{2}} \sum i^{L-L'}(-)^{-J-\frac{1}{2}}[J]^2 
  \begin{Bmatrix}
  \frac{3}{2} & L & J \\   L' & \frac{1}{2} & 2
  \end{Bmatrix}
M^{J\frac{3}{2}}_{L'L}(q,k) a'_{L'L} \nonumber \\
{\cal G}_5 & = & \sqrt{\frac{1}{2}} \sum i^{L-L'}(-)^{-J-\frac{1}{2}}[J]^2 
  \begin{Bmatrix}
  \frac{3}{2} & L & J \\   L' & \frac{1}{2} & 2
  \end{Bmatrix}
M^{J\frac{3}{2}}_{L'L}(q,k) b'_{L'L} \nonumber \\
{\cal G}_6 & = & \frac{1}{4\pi} \left(\frac{1+\sqrt{2}}{\sqrt{6}}\right) 
\sum i^{L-L'+1}(-)^{-J-\frac{1}{2}}[LL'][J]^2 (L0L'0|10)
  \begin{Bmatrix}
  S & L & J \\   L' & \frac{1}{2} & 1
  \end{Bmatrix}
\nonumber \\
& \times & 
M^{JS}_{L'L}(q,k) P_{L'}(\hat k \cdot \hat q) \ .
\label{SSTRUC12na}
\end{eqnarray}
The summations are over all quantum numbers appearing explicitly in the r.h.s. of
the equalities.

The above result is the most general form of the spin structure of the transition
operator for a negative parity baryon $B$. As in the positive parity case, we 
note that the amplitude in Eq.(\ref{SSTRUC12n}) has six independent spin operators 
and is valid for both real and virtual photons. Using Eq.(\ref{tensor}), it can be
rewritten as
\begin{eqnarray}
\hat M^-(\vec q, \vec k) & = & iG_1\ \vec\epsilon \cdot \hat q + 
G_2\ \vec\sigma \cdot (\vec\epsilon \times \hat q) + 
G_3\ \vec\sigma \cdot (\vec\epsilon \times \hat k) \nonumber \\
& + &
G_4\ [\vec\sigma \cdot \hat k \vec\epsilon \cdot \hat n_2 +
     \vec\sigma \cdot \hat n_2 \vec\epsilon \cdot \hat k ] + 
G_5\ [\vec\sigma \cdot \hat q \vec\epsilon \cdot \hat n_2 +
     \vec\sigma \cdot \hat n_2 \vec\epsilon \cdot \hat q ] +
iG_6\ \vec\epsilon \cdot \hat k \ ,
\label{SSTRUC_1N}
\end{eqnarray}
where 
\begin{eqnarray}
G_1 & = &  \frac{1}{|\hat k \times \hat q|} {\cal G}_1 \ , \ \ 
G_2  =  \frac{1}{|\hat k \times \hat q|} {\cal G}_2 \ , \ \
G_3  =  {\cal G}_3 - (\hat k \cdot \hat q) G_2 \ , \nonumber \\
G_4 & = & \frac{1}{2}{\cal G}_4 \ , \ \ \ \ \ \ \ \ \ 
G_5  =  \frac{1}{2}{\cal G}_5 \ , \ \ \ \ \ \ \ \ \
G_6  =  {\cal G}_6 - (\hat k \cdot \hat q) G_2 \ .
\label{SSTRUC_1Pan}
\end{eqnarray}

Quite recently, Zhao and Al-Khalili \cite{Zhao} have also given the spin 
structure of the photoproduction amplitude in connection to the reaction 
$\gamma N \to 
\bar K \Theta^+$ for the case of negative parity $\Theta^+$. The structure given 
in Eq.(\ref{SSTRUC_1N}) with $\vec\epsilon\cdot\hat k=0$ 
is equivalent to that of Eq.(18) in Ref.\cite{Zhao}, except for the term 
$\vec\sigma \cdot \hat n_2 \vec\epsilon \cdot \hat q$  which 
has not been included in Ref.\cite{Zhao} on the grounds that it is a higher-order 
contribution. However, this term and the $\vec\sigma \cdot \hat q \vec\epsilon 
\cdot \hat n_2$ term contribute with the same coefficient $G_5$. We also note that 
the recent model-independent analysis of the $\Theta^+$ photoproduction 
\cite{Nak5} has been based on the present results and, in particular, on 
Eqs.(\ref{SSTRUC12_1P},\ref{SSTRUC_1N}) with $\vec\epsilon\cdot\hat k=0$ .

\section{Application: near threshold amplitude in $\gamma N \to M B$.}

As an application of the present results, we consider the reaction
$\gamma N \to M B$ in the near-thresholds kinematics. The complete transition 
amplitude is given by Eq.(\ref{SSTRUC12_1P}) for a positive parity baryon $B$ 
and by Eq.(\ref{SSTRUC_1N}) for a negative baryon $B$, respectively, with 
$\vec\epsilon\cdot\hat k=0$. In the near-threshold energy region, 
the final $M B$ is mainly in relative $S$ and $P$ waves. Then, 
considering only $L'=0,1$, there are seven partial-wave amplitudes. For positive 
parity $B$, they are
\begin{equation}
^1S_1  \to  S_1 \ , \ \ \ \ ^3D_1  \to  S_1 \ , \ \ \ \ 
^{3,1}P_1 \to P_1 \ , \ \ \ \ ^{3,1}P_3  \to  P_3 \ , \ \ \ \ ^3F_3 \to P_3 \ , 
\label{SPpw_p1}
\end{equation}
where we have used the notation $^{2S}L_{2J} \to L'_{2J}$. For these amplitudes, 
the coefficients $F_i, (i=1,...,4)$ in Eq.(\ref{SSTRUC12_1P}) exhibit the following 
angular and energy (due to $q^{L'}$) dependences
\begin{equation}
F_1  =   A_0 + \left(\frac{q}{\Lambda}\right) A_1\cos(\theta) \ , \ \ \ \
F_2  =   \left(\frac{q}{\Lambda}\right) B_1\sin(\theta) \ , \ \ \ \
F_3  =   \left(\frac{q}{\Lambda}\right) C_1 \ , \ \ \ \ 
F_4  =   0 \ .
\label{SSTRUC1_P_thr}
\end{equation}
In the above equation, $A_{L'}$, $B_{L'}$ and $C_{L'}$ denote the linear 
combinations of the partial-wave matrix elements $M^{JS}_{L'L}$ resulting from 
Eqs.(\ref{SSTRUC12a},\ref{SSTRUC12_1Pa}) for those states specified in 
Eq.(\ref{SPpw_p1}) having orbital angular momentum $L'$. They are given explicitly 
in Appendix B. Note that the factor $(q/\Lambda)^{L'}$ contained in $M^{JS}_{L'L}$ 
due to the centrifugal barrier has been displayed explicitly in the above equation. 
$\Lambda$ is a typical scale of the problem which may be taken to be the 
four-momentum transfer at threshold, 
$\Lambda^2 \sim -t = [m_B - m_N^2/(m_B+m_M)]m_M$,  
where $m_B$, $m_M$ and $m_N$ denote the masses of the 
baryon $B$, meson $M$ and nucleon, respectively. Therefore, near threshold, 
higher partial-wave contributions will be suppressed by the factor $(q/\Lambda)^{L'}$
if heavy particles are produced in the final state. Moreover, 
for short-range processes, the coefficients $A_{L'}$, 
$B_{L'}$ and $C_{L'}$ are nearly constant independent of energy. 

Analogously, for a negative parity baryon $B$, the possible partial-wave amplitudes
are
\begin{equation}
^{3,1}P_1  \to  S_1 \ , \ \ \ \ ^1S_1  \to  P_1 \ , \ \ \ \ 
^3D_1 \to P_1 \ , \ \ \ \ ^3S_3  \to  P_3 \ , \ \ \ \ ^{3,1}D_3 \to P_3 \ . 
\label{SPpw_n1}
\end{equation}
With these amplitudes, the coefficients $G_i$ in Eq.(\ref{SSTRUC_1N}) exhibit the 
following angular and energy (due to $q^{L'}$) dependences
\begin{eqnarray}
G_1 & = &  \left(\frac{q}{\Lambda}\right) A'_1 \ , \ \ \ \ \ \ \ \ \ \ \ \ \ \ \ \ \ \ \ \
G_2  =   \frac{1}{\sqrt{2}} \left( \frac{\sqrt{2}-1}{\sqrt{2}+1} \right) G_1 \ , \nonumber \\
G_3 & = &  B'_0 +  \left(\frac{q}{\Lambda}\right) B'_1 \cos(\theta) \ , \ \ \ \ 
G_4  =   \left(\frac{q}{\Lambda}\right) C'_1 \sin(\theta) \ , \ \ \ \ 
G_5  = 0 \ .  
\label{SSTRUC1_N_thr}
\end{eqnarray}
In the above equation, $A'_{L'}$, $B'_{L'}$ and $C'_{L'}$ denote the linear 
combinations of the partial-wave matrix elements $M^{JS}_{L'L}$ resulting from 
Eqs.(\ref{SSTRUC12na},\ref{SSTRUC_1Pan}) for those states specified in 
Eq.(\ref{SPpw_n1}) having orbital angular momentum $L'$. They are given explicitly 
in Appendix B. For a short range process, $A'_{L'}$, $B'_{L'}$ and $C'_{L'}$ are 
nearly constant independent of energy.

Following Ref.\cite{Nak5} we now introduce $\vec\epsilon_\perp \equiv \hat y$ and 
$\vec\epsilon_\parallel \equiv \hat x$ denoting the photon polarization perpendicular 
to and lying in the reaction plane ($xz$-plane), respectively. Recall that the 
reaction plane is defined as the plane containing the vectors $\vec k$ 
(in the $+z$-direction) and $\vec q$ and that $\vec k \times \vec q$ is along the 
$+y$-direction, in which case, $\hat n_1 = \hat x$ and $\hat n_2 = \hat y$. 
Then, from Eqs.(\ref{SSTRUC12_1P},\ref{SSTRUC1_P_thr})
\begin{eqnarray}
\hat M^{+\perp} & = & \left[A_0 + \left( \frac{q}{\Lambda}\right) A_1\cos(\theta)\right] \sigma_y  
                  + i \left( \frac{q}{\Lambda}\right) B_1 \sin(\theta) \ , \nonumber \\
\hat M^{+\parallel} & = & \left[A_0 + \left( \frac{q}{\Lambda}\right) A_1\cos(\theta)\right] \sigma_x  
+ \ \ \left( \frac{q}{\Lambda}\right) C_1 \sin(\theta)\ \sigma_z  \ .
\label{PPOLM+} 
\end{eqnarray}

Similarly, from Eqs.(\ref{SSTRUC_1N},\ref{SSTRUC1_N_thr})
\begin{eqnarray}
\hat M^{-\perp} & = & \ \ \ \left[B'_0 +  \left(\frac{q}{\Lambda}\right) \bar B'_1 \cos(\theta)\right] \sigma_x  
+ \ \ \left(\frac{q}{\Lambda}\right) \bar C'_1 \sin(\theta) \sigma_z \ , \nonumber \\
\hat M^{-\parallel} & = & - \left[B'_0 +  \left(\frac{q}{\Lambda}\right) \bar B'_1 \cos(\theta)\right] \sigma_y  
+ i\left(\frac{q}{\Lambda}\right) A'_1 \sin(\theta) \ .
\label{PPOLM-} 
\end{eqnarray}
In the above equation, $\bar B'_1 \equiv B'_1 + \bar A'_1$  and 
$\bar C'_1 \equiv C'_1 - \bar A'_1$, with $\bar A'_1 \equiv 
\frac{1}{\sqrt{2}} \left( \frac{\sqrt{2}-1}{\sqrt{2}+1} \right) A'_{1}$.

Following Ref.\cite{Nak5}, any observable in the $\gamma N \to M B$ 
reaction can be readily calculated from Eqs.(\ref{PPOLM+},\ref{PPOLM-}) 
for both positive and negative parity baryons $B$. It is then straightforward 
to show that no observable in this reaction is able to distinguish between 
positive and negative parity $B$, unless one measures the polarization 
of $B$. In particular, neither the energy 
dependence nor the angular distribution exhibits features sufficiently distinct
to determine the parity of the baryon $B$ unambiguously. As has been shown in 
Ref.\cite{HHNM} in connection with $NN \to Y\Theta^+$, 
the situation is quite different in the 
$NN \to B' B$ reaction. We shall discuss this latter reaction in the 
following two sections.

\section{The reaction $N + N \to B' + B$.}

We now consider the process $N + N \to B' + B$, where $B'$ stands for a positive parity 
spin-1/2 baryon and $B$, either a positive or negative parity spin-1/2 baryon.
The partial-wave expansion of the corresponding reaction amplitude is
\begin{eqnarray}
<S'M_{S'}|\hat M(\vec p\ ', \vec p)|SM_S>  = 
& \sum & i^{L-L'} (S'M_{S'}L'M_{L'}|JM_{J})
(SM_SLM_L|JM_J) M^{S'SJ}_{L'L}(p', p) \nonumber \\ 
& \times & Y_{L'M_{L'}}(\hat p') Y^*_{LM_L}(\hat p) \ , 
\label{PWE}
\end{eqnarray}
where $S, L, J$ stand for the total spin, total orbital angular momentum and the total 
angular momentum, respectively, of the initial $NN$ state. $M_S$, $M_L$ and $M_J$ denote 
the corresponding projection quantum numbers. The primed quantities stand for the 
corresponding quantum numbers of the final $B'B$ state. The summation runs 
over all quantum numbers not specified in the l.h.s. of Eq.(\ref{PWE}). 
$\vec p$ and $\vec p\ '$ denote the relative momenta of the two particles in the initial 
and final states, respectively. 
We note that, in Eq.(\ref{PWE}), apart from the restrictions on the quantum numbers encoded 
in the geometrical factors, total parity conservation imposes that $(-)^{L+L'} = +1$ and
$(-)^{L+L'} = -1$ for positive and negative parity $B$, respectively.

Eq.(\ref{PWE}) can be inverted to solve for the partial-wave matrix element
$M^{S'SJ}_{L'L}(p', p)$. We have
\begin{eqnarray}
M^{S'SJ}_{L'L}(p', p) & = &
\sum i^{L'-L} (S'M_{S'}L'M_{L'}|JM_{J})(SM_SL0|JM_J) \frac{8\pi^2}{2J+1} \nonumber \\
& \times & \sqrt{\frac{2L+1}{4\pi}}
\int_{-1}^{+1} d(\cos(\theta)) Y_{L'M_{L'}}^*(\theta, 0)
<S'M_{S'}|\hat M(\vec p\ ', \vec p)|SM_S> \ ,
\label{pwme}
\end{eqnarray}
where, without loss of generality, the $z$-axis is chosen along $\vec p$ and  
$\vec p\ '$ in the $xz$-plane; $\cos(\theta) = \hat p' \cdot \hat p$. The summation 
is over all quantum numbers not specified in the l.h.s. of the equation.

The most general spin structure of the transition operator amplitude can be extracted from 
Eq.(\ref{PWE}) as
\begin{equation}
\hat M(\vec p\ ', \vec p) = \sum_{S'SM_{S}M_{S'}} 
|S'M_{S'}> <S'M_{S'}|\hat M(\vec p\ ', \vec p)|SM_S> <SM_S| \ .
\label{SSTRUC0}
\end{equation}
Inserting Eq.(\ref{PWE}) into Eq.(\ref{SSTRUC0}) and re-coupling gives
\begin{eqnarray}
\hat M(\vec p\ ', \vec p) & = & \sum i^{L-L'} (-)^{J+S'} [J]^2  
M^{S'SJ}_{L'L}(p', p) \sum_{\alpha}
  \begin{Bmatrix}
  S & L & J \\   L' & S' & \alpha
  \end{Bmatrix} \nonumber \\ 
& \times &
[A_{S'} \otimes B_S]^\alpha \cdot [Y_{L'}(\hat p') \otimes Y_L(\hat p)]^\alpha \ ,
\label{SSTRUC1}
\end{eqnarray}
where we have used the notations $B_{SM_{S}}\equiv (-)^{S-M_S}<S-M_S|$ and
$A_{S'M_{S'}}\equiv |S'M_{S'}>$.

We now expand $[A_{S'} \otimes B_S]^\alpha$, for each tensor of rank $\alpha$, 
in terms of the complete set of available spin operators in the problem, i.e., the 
Pauli spin matrices $\vec \sigma_1$ and $\vec \sigma_2$, corresponding to the interacting
particles 1 and 2, together with the identity matrix. Then, $\alpha$ takes the values 
0, 1, and 2, and denotes the rank of the corresponding (spin) tensor. There are six cases 
to be considered:
\begin{eqnarray}      
S=S'=0, \alpha=0: \ \ \ \ \ \ \ \ 
[A_{S'} \otimes B_S]^0 & = & |00><00|\equiv P_{S=0}  \ ,\nonumber \\
S=S'=1, \alpha=0: \ \ \ \ \ \ \ \
[A_{S'} \otimes B_S]^0 & = & \frac{1}{\sqrt{3}}\sum_{M_S}|1M_S><1M_S|
\equiv \frac{1}{\sqrt{3}} P_{S=1}  \ ,\nonumber \\
S=S'=1, \alpha=1: \ \ \ \ \ \ \ \
[A_{S'} \otimes B_S]^1 & = & \frac{1}{2\sqrt{2}} 
\left( \vec\sigma_1 + \vec\sigma_2 \right)  \ , \nonumber \\
S=0, S'=1, \alpha=1: \ \ \ \ \ \ \ \ 
[A_{S'} \otimes B_S]^1 & = & \frac{1}{2}  
\left( \vec\sigma_1 - \vec\sigma_2 \right) P_{S=0} \ ,
\ \ \ \ \ \ \ \ \  \ \ \ \ \ \ \ \ \ \ \ \ \ \ \ \ \  \nonumber \\
S=1, S'=0, \alpha=1: \ \ \ \ \ \ \  \
[A_{S'} \otimes B_S]^1 & = & - \frac{1}{2}  
\left( \vec\sigma_1 - \vec\sigma_2 \right) P_{S=1} \ ,
\ \ \ \ \ \ \ \ \  \ \ \ \ \ \ \ \ \ \ \ \ \  \ \ \ \ \nonumber \\
S=S'=1, \alpha=2: \ \ \ \ \ \ \ \
[A_{S'} \otimes B_S]^2 & = & \frac{1}{\sqrt{2}} [\vec\sigma_1 \otimes \vec\sigma_2]^2  \ ,
\label{SOPER}
\end{eqnarray}
where $P_{S}$ stands for the spin projection operator onto the (initial) spin 
singlet and triplet states as $S=0$ and $1$, respectively. In terms of the Pauli spin 
matrices we have 
$P_{S=0} = (1 - \vec\sigma_1\cdot\vec\sigma_2) / 4$ and 
$P_{S=1} = (3 + \vec\sigma_1\cdot\vec\sigma_2) / 4$.
Also, 
$(\vec\sigma_1 - \vec\sigma_2)P_S/2 = [(\vec\sigma_1 - \vec\sigma_2) +
(-)^S i (\vec\sigma_1 \times \vec\sigma_2)]/4$.

In the following we shall consider the case of a positive and a negative parity baryon $B$ 
separately.

\subsection{Positive parity case}

For a positive parity baryon $B$, the quantity $[Y_{L'}(\hat p') \otimes Y_L(\hat p)]^\alpha$ 
in Eq.(\ref{SSTRUC1}) can be read off from Eq.(\ref{aux11}). Note that 
$[Y_{L'}(\hat p') \otimes Y_L(\hat p)]^\alpha = 
(-)^\alpha [Y_L(\hat p) \otimes Y_{L'}(\hat p')]^\alpha$ for the positive parity case. 
Inserting this and Eq.(\ref{SOPER}) into Eq.(\ref{SSTRUC1}), we have
\begin{eqnarray}
\hat M^+(\vec p\ ', \vec p) & = & {\cal D}_1 P_{S=0} + {\cal D}_2 P_{S=1} \nonumber \\
& + & i{\cal D}_3 (\vec\sigma_1 + \vec\sigma_2)\cdot \hat n_2
+ i{\cal D}_4 (\vec\sigma_1 - \vec\sigma_2)\cdot \hat n_2 P_{S=0}
+ i{\cal D}_5 (\vec\sigma_1 - \vec\sigma_2)\cdot \hat n_2 P_{S=1} \nonumber \\
& + & {\cal D}_6 [\vec\sigma_1 \otimes \vec\sigma_2]^2 \cdot [\hat p \otimes \hat p]^2
+ {\cal D}_7 [\vec\sigma_1 \otimes \vec\sigma_2]^2 \cdot [\hat p' \otimes \hat p']^2
+ {\cal D}_8 [\vec\sigma_1 \otimes \vec\sigma_2]^2 \cdot [\hat p \otimes \hat p']^2 \ ,  
\nonumber \\
\label{SSTRUC1p}
\end{eqnarray}
where
\begin{eqnarray}
{\cal D}_1 & = & 
\frac{1}{4\pi} \sum [L]^2 M^{00L}_{LL}(p', p) P_L(\hat p' \cdot \hat p) \ , \nonumber \\
{\cal D}_2 & = & 
\frac{1}{4\pi} \frac{1}{3} \sum [J]^2 M^{11J}_{LL}(p', p) P_L(\hat p' \cdot \hat p)
 \ , \nonumber \\
{\cal D}_3 & = & 
- \frac{1}{4\pi} \frac{1}{8} \sum [J]^2 \left( 1 + \frac{2 - J(J+1)}{L(L+1)}\right)
M^{11J}_{LL}(p', p) P^1_L(\hat p' \cdot \hat p) \ , \nonumber \\
{\cal D}_4 & = &
- \frac{1}{4\pi} \frac{1}{2} \sum \frac{[L]^2 }{\sqrt{L(L+1)}}
M^{10L}_{LL}(p', p) P^1_L(\hat p' \cdot \hat p) \ , \nonumber \\
{\cal D}_5 & = &
  \frac{1}{4\pi} \frac{1}{2} \sum \frac{[L]^2}{\sqrt{L(L+1)}}
M^{01L}_{LL}(p', p) P^1_L(\hat p' \cdot \hat p) \ , \nonumber \\
{\cal D}_6 & = & 
\frac{1}{2} \sum i^{L-L'} (-)^{J+1} [J]^2 
  \begin{Bmatrix}
  1 & L & J \\   L' & 1 & 2
  \end{Bmatrix}
M^{11J}_{L'L}(p', p) a_{L'L} \ , \nonumber \\
{\cal D}_7 & = & 
\frac{1}{2} \sum i^{L-L'} (-)^{J+1} [J]^2 
  \begin{Bmatrix}
  1 & L & J \\   L' & 1 & 2
  \end{Bmatrix}
M^{11J}_{L'L}(p', p) b_{L'L} \ , \nonumber \\
{\cal D}_8 & = & 
\frac{1}{2} \sum i^{L-L'} (-)^{J+1} [J]^2 
  \begin{Bmatrix}
  1 & L & J \\   L' & 1 & 2
  \end{Bmatrix}
M^{11J}_{L'L}(p', p) c_{L'L} \ , 
\label{SSTRUC1_pam}
\end{eqnarray}
The coefficients $a_{L'L}$, $b_{L'L}$ and $c_{L'L}$ are given by 
Eqs.(\ref{a4},\ref{a5}) with the replacements $\hat k \to \hat p$ and $\hat q \to \hat p '$;
these same replacements are needed to calculate $\hat n_2$ in Eq.(\ref{SSTRUC1p}).

Eq.(\ref{SSTRUC1p}) is the most general spin structure of the transition operator for
a positive parity baryon $B$, 
consistent with symmetry principles. It contains eight independent spin structures. 
The first two terms are the central spin singlet and triplet interactions, respectively. 
The third term is the spin-orbit interaction. The fourth and fifth terms describe the 
spin singlet$\to$triplet and triplet$\to$singlet transitions, respectively. The last three 
terms are the tensor interactions of rank 2. Apart from the fourth and fifth terms, all 
the other terms conserve the total spin in the transition.

It should be mentioned that in the case of identical particles, i.e., $NN \to NN$, the
structure $(\vec\sigma_1 - \vec\sigma_2)$ is not allowed in the scattering amplitude, which 
reduces the total number of independent spin structures in Eq.(\ref{SSTRUC1p}) to six
\cite{NL}. Furthermore, for elastic scattering, ${\cal D}_6 = {\cal D}_7$, as a consequence 
of time reversal invariance \cite{NL}.

For the purpose of calculating the observables directly from the transition operator of
Eq.(\ref{SSTRUC1p}), it is convenient to re-express it in the form 
\begin{equation}
\hat M^+ = \sum_{\lambda=1}^9 \sum_{n,n'=0}^3 
M^\lambda_{nn'} \sigma_n(1) \sigma_{n'}(2) \ ,
\label{ampl_obsp}
\end{equation}
where $\sigma_0(i) \equiv 1$, $\sigma_1(i) \equiv \sigma_x(i)$, etc., for $i$-th nucleon. 
The coefficients $M^\lambda_{nn'}$ are linear combinations of the coefficients appearing in 
Eq.(\ref{SSTRUC1_pam}). Explicitly, we have
\begin{eqnarray}
M^1_{nn'} & = & \frac{1}{4} \left[3{\cal D}_2 + {\cal D}_1\right] \delta_{n,0} \delta_{n',0} \ , \nonumber \\
M^2_{nn'} & = & i\left[{\cal D}_3 + \frac{1}{2}({\cal D}_4 + {\cal D}_5) \right] \hat n_{2n} (1-\delta_{n,0}) \delta_{n',0} \ , \nonumber \\
M^3_{nn'} & = & i\left[{\cal D}_3 - \frac{1}{2}({\cal D}_4 + {\cal D}_5) \right] \hat n_{2n'} \delta_{n,0} (1-\delta_{n',0}) \ , \nonumber \\
M^4_{nn'} & = &  \frac{1}{2}\left[{\cal D}_5 - {\cal D}_4 \right] \varepsilon_{nn'k} \hat n_{2k} 
                 (1-\delta_{n,0})(1- \delta_{n',0}) \ , \nonumber \\
M^5_{nn'} & = &  \left[\frac{1}{4}({\cal D}_2 - {\cal D}_1) - \frac{1}{3}({\cal D}_6 + {\cal D}_7  + 
                 (\hat p \cdot \hat p'){\cal D}_8) \right] \delta_{n,n'}(1-\delta_{n,0}) \ , \nonumber \\
M^6_{nn'} & = &  {\cal D}_6 \hat p_n \hat p_{n'} (1-\delta_{n,0}) (1-\delta_{n',0}) \ , \nonumber \\
M^7_{nn'} & = &  {\cal D}_7 \hat p'_n \hat p'_{n'} (1-\delta_{n,0}) (1-\delta_{n',0}) \ , \nonumber \\
M^8_{nn'} & = &  \frac{1}{2}{\cal D}_8 \hat p_n \hat p'_{n'} (1-\delta_{n,0}) (1-\delta_{n',0}) \ , \nonumber \\
M^9_{nn'} & = &  \frac{1}{2}{\cal D}_8 \hat p'_n \hat p_{n'} (1-\delta_{n,0}) (1-\delta_{n',0}) \ ,
\label{ampl1_obsp}
\end{eqnarray}
where we have used the notation $\hat a_n$ for the $n$-th component of an arbitrary unit vector $\hat a$; 
$\varepsilon_{nn'k}$ denotes the antisymmetric Levi-Civita tensor and $\lambda$ is an 
index for the type of term (operator) being considered. 
In obtaining Eq.(\ref{ampl1_obsp}), we have made use of the identity 
\begin{equation}
3 [\ \vec\sigma_1 \otimes \vec\sigma_2\ ]^2 \cdot [\ \hat a \otimes 
\hat b\ ]^2 = \frac{3}{2} [\ \vec\sigma_1 \cdot \hat a \vec\sigma_2 \cdot \hat b + \vec\sigma_1 \cdot 
\hat b \vec\sigma_2 \cdot \hat a \ ] - (\hat a \cdot \hat b) \vec\sigma_1 \cdot \vec\sigma_2 \ ,
\label{tensor_NN}
\end{equation}
where $\hat a$ and $\hat b$ denote arbitrary unit vectors.

With the transition operator in the form of Eq.(\ref{ampl_obsp}), it is straightforward 
to calculate any observable of interest. Of course, it can also be expressed in terms 
of the matrix elements of total spin of Eq.(\ref{PWE}); for convenience some observables 
are given in Appendix D in both representations.     

\subsection{Negative parity case}

For a negative parity baryon $B$, the quantity $[Y_{L'}(\hat p') \otimes Y_L(\hat p)]^\alpha$ 
in Eq.(\ref{SSTRUC1}) can be read off from Eq.(\ref{aux11n}). For this case
$[Y_{L'}(\hat p') \otimes Y_L(\hat p)]^\alpha = (-)^{\alpha + 1} [Y_L(\hat p) \otimes Y_{L'}(\hat p')]^\alpha$.
Inserting this and Eq.(\ref{SOPER}) into Eq.(\ref{SSTRUC1}), we have
\begin{eqnarray}
\hat M^-(\vec p\ ', \vec p) & = & 
i ({\cal H}_1 \hat n_1 + {\cal H}_2 \hat p ) \cdot (\vec\sigma_1 + \vec\sigma_2)  \nonumber \\
& + & i ({\cal H}_3 \hat n_1 + {\cal H}_4 \hat p ) \cdot (\vec\sigma_1 - \vec\sigma_2) P_{S=0} 
 +  i ({\cal H}_5 \hat n_1 + {\cal H}_6 \hat p ) \cdot (\vec\sigma_1 - \vec\sigma_2) P_{S=1} \nonumber \\
& + & {\cal H}_7 [\vec\sigma_1 \otimes \vec\sigma_2]^2 \cdot [\hat p \otimes \hat n_2]^2
+ {\cal H}_8 [\vec\sigma_1 \otimes \vec\sigma_2]^2 \cdot [\hat p' \otimes \hat n_2]^2 \ ,
\label{SSTRUC1n}
\end{eqnarray}
where
\begin{eqnarray}
{\cal H}_1 & = & 
\frac{1}{4\pi} \frac{1}{2} \sum i^{L-L'+1} (-)^J [J]^2 \frac{[LL']}{\sqrt{L'(L'+1)}}  
(L0L'1|11) 
  \begin{Bmatrix}
  1 & L & J \\   L' & 1 & 1
  \end{Bmatrix}
M^{11J}_{L'L}(p', p) P^1_{L'}(\hat p' \cdot \hat p) \ , \nonumber \\
{\cal H}_2 & = & 
\frac{1}{4\pi} \frac{1}{2\sqrt{2}} \sum i^{L-L'+1} (-)^J [LL'][J]^2 
(L0L'0|10) 
  \begin{Bmatrix}
  1 & L & J \\   L' & 1 & 1
  \end{Bmatrix}
M^{11J}_{L'L}(p', p) P_{L'}(\hat p' \cdot \hat p) \ , \nonumber \\
{\cal H}_3 & = & 
\frac{1}{4\pi} \frac{1}{\sqrt{6}} \sum i^{L-L'+1} (-)^{L'+1} \frac{[L'][L]^2}{\sqrt{L'(L'+1)}}  
(L0L'1|11) 
M^{10L}_{L'L}(p', p) P^1_{L'}(\hat p' \cdot \hat p) \ , \nonumber \\
{\cal H}_4 & = & 
\frac{1}{4\pi} \frac{1}{2\sqrt{3}} \sum i^{L-L'+1} (-)^{L'+1}[L'][L]^2 
(L0L'0|10) 
M^{10L}_{L'L}(p', p) P_{L'}(\hat p' \cdot \hat p) \ , \nonumber \\
{\cal H}_5 & = & 
\frac{1}{4\pi} \frac{1}{\sqrt{6}} \sum i^{L-L'+1} (-)^{L} \frac{[L][L']^2}{\sqrt{L'(L'+1)}}  
(L0L'1|11) 
M^{01L'}_{L'L}(p', p) P^1_{L'}(\hat p' \cdot \hat p) \ , \nonumber \\
{\cal H}_6 & = & 
\frac{1}{4\pi} \frac{1}{2\sqrt{3}} \sum i^{L-L'+1} (-)^{L}[L][L']^2 
(L0L'0|10) 
M^{01L'}_{L'L}(p', p) P_{L'}(\hat p' \cdot \hat p) \ , \nonumber \\
{\cal H}_7 & = & 
\frac{1}{2} \sum i^{L-L'} (-)^J [J]^2 
  \begin{Bmatrix}
  1 & L & J \\   L' & 1 & 2
  \end{Bmatrix}
M^{11J}_{L'L}(p', p) a'_{L'L} \ , \nonumber \\
{\cal H}_8 & = & 
\frac{1}{2} \sum i^{L-L'} (-)^J [J]^2 
  \begin{Bmatrix}
  1 & L & J \\   L' & 1 & 2
  \end{Bmatrix}
M^{11J}_{L'L}(p', p) b'_{L'L} \ ,
\label{SSTRUC1_nam}
\end{eqnarray}
The coefficients $a'_{L'L}$ and  $b'_{L'L}$ are given by 
Eqs.(\ref{a9},\ref{a10}) with the replacements $\hat k \to \hat p$ and $\hat q \to \hat p '$.
The same replacement is also required to calculate $\hat n_2$. 

Eq.(\ref{SSTRUC1n}) is the most general spin structure of the transition operator for
a negative parity baryon $B$, 
consistent with symmetry principles. It also contains eight independent spin structures, 
but no central interaction is present in this case.

Analogous to the positive parity case, Eq.(\ref{SSTRUC1n}) can be re-expressed in the form  
\begin{equation}
\hat M^- = \sum_{\lambda=1}^{10} \sum_{n,n'=0}^3 
M^\lambda_{nn'} \sigma_n(1) \sigma_{n'}(2) \ ,
\label{ampl_obsn}
\end{equation}
where the coefficients $M^\lambda_{nn'}$ are given by 
\begin{eqnarray}
M^1_{nn'} & = & i\left[{\cal H}_1 + \frac{1}{2}({\cal H}_3 + {\cal H}_5) \right] \hat n_{1n} (1-\delta_{n,0}) \delta_{n',0} \ , \nonumber \\
M^2_{nn'} & = & i\left[{\cal H}_2 + \frac{1}{2}({\cal H}_4 + {\cal H}_6) \right] \hat p_n    (1-\delta_{n,0}) \delta_{n',0} \ , \nonumber \\
M^3_{nn'} & = & i\left[{\cal H}_1 - \frac{1}{2}({\cal H}_3 + {\cal H}_5) \right] \hat n_{1n'} \delta_{n,0} (1-\delta_{n',0}) \ , \nonumber \\
M^4_{nn'} & = & i\left[{\cal H}_2 - \frac{1}{2}({\cal H}_4 + {\cal H}_6) \right] \hat p_{n'}  \delta_{n,0} (1-\delta_{n',0}) \ , \nonumber \\
M^5_{nn'} & = & i \frac{1}{2}\left[{\cal H}_3 - {\cal H}_5 \right] \varepsilon_{nn'k} \hat n_{1k} 
                 (1-\delta_{n,0})(1- \delta_{n',0}) \ , \nonumber \\
M^6_{nn'} & = & i \frac{1}{2}\left[{\cal H}_4 - {\cal H}_6 \right] \varepsilon_{nn'k} \hat p_k 
                 (1-\delta_{n,0})(1- \delta_{n',0}) \ , \nonumber \\
M^7_{nn'} & = &  \frac{1}{2}{\cal H}_7 \hat p_n \hat n_{2n'} (1-\delta_{n,0}) (1-\delta_{n',0}) \ , \nonumber \\
M^8_{nn'} & = &  \frac{1}{2}{\cal H}_7 \hat p_{n'} \hat n_{2n} (1-\delta_{n,0}) (1-\delta_{n',0}) \ , \nonumber \\
M^9_{nn'} & = &  \frac{1}{2}{\cal H}_8 \hat p'_n \hat n_{2n'} (1-\delta_{n,0}) (1-\delta_{n',0}) \ , \nonumber \\
M^{10}_{nn'} & = &  \frac{1}{2}{\cal H}_8 \hat p'_{n'} \hat n_{2n} (1-\delta_{n,0}) (1-\delta_{n',0}) \ .
\label{ampl1_obsn}
\end{eqnarray}
\section{Application: near threshold amplitude in $p p \to B' B$.}
As an application of the results of the previous section, we consider the
reaction $p p \to B' B$ in the near-threshold energy region. We restrict 
to $S$ and $P$ waves in the final state, i.e., $L'=0,1$. In contrast to the 
photoproduction reaction discussed in section II, here the Pauli principle restricts the
initial $pp$ state to $(-)^{S+L}=+1$.

For a positive parity baryon $B$, we then have six partial-wave states
\begin{equation}
^1S_0 \to\,^1S_0 \ , \ \ \ \ ^3P_1 \to\,^1P_1 \ , \ \ \ \  ^3P_{0,1,2} \to\,^3P_{0,1,2} \ , \ \ \ \ 
^3F_2 \to\,^3P_2 \ .
\label{pw_p}
\end{equation}
Here we use the notation $^{2S+1}L_J \to\,^{2S'+1}L'_J$. Then, the coefficients 
${\cal D}_i$ in Eq.(\ref{SSTRUC1_pam}) reduce to
\begin{eqnarray}
{\cal D}_1 & = & A^0_0 \ , \ \ \ \
{\cal D}_2  =  \left(\frac{p'}{\Lambda}\right) A^1_1 \cos(\theta) \ , \ \ \ \ 
{\cal D}_3  =  \left(\frac{p'}{\Lambda}\right) B^1_1 \sin(\theta) \ , \ \ \ \ 
{\cal D}_4  =  0 \ , \nonumber \\
{\cal D}_5 & = & \left(\frac{p'}{\Lambda}\right) C^1_1 \sin(\theta) \ , \ \ \ \ \ \
{\cal D}_6  =  0 \ , \ \ \ \
{\cal D}_7  =  \left(\frac{p'}{\Lambda}\right) D^1_1 \cos(\theta) \ , \ \ \ \
{\cal D}_8  =  \left(\frac{p'}{\Lambda}\right) E^1_1 \ , 
\label{SSTRUC1_pam_pw}
\end{eqnarray}
where the quantities $A^S_{L'}$, $B^S_{L'}$, etc, are given in Appendix C.
Here, again, the $(p'/\Lambda)^{L'}$ dependence of the partial-wave matrix 
elements due to the centrifugal barrier is displayed explicitly. As in the 
photoproduction reaction discussed in section II, the scale $\Lambda$ may 
be taken to be the four-momentum transfer at threshold, 
$\Lambda^2 \sim -t = m_B m_{B'} - m_N^2$, where
$m_B$ denotes the mass of the baryon $B$ and $m_{B'}$, the mass of the 
positive parity baryon $B'$.
  
For a negative parity baryon $B$, the possible partial-wave transitions are
\begin{equation}
^3P_0 \to\,^1S_0 \ , \ \ \ \  ^3P_1 \to\,^3S_1 \ , \ \ \ \  ^1S_0 \to\,^3P_0 \ ,
\ \ \ \  ^1D_2 \to\,^3P_2 \ .
\label{pw_n}
\end{equation}
With these partial-wave states, Eq.(\ref{SSTRUC1_nam}) reduces to
\begin{eqnarray}
{\cal H}_1 & = &  0  \ , \ \ \ \ 
{\cal H}_2  =  A^{'11}_0  \ , \ \ \ \ 
{\cal H}_3  =  \left(\frac{p'}{\Lambda}\right) B^{'10}_1 \sin(\theta) \ , \ \ \ \   
{\cal H}_4  =  \left(\frac{p'}{\Lambda}\right) C^{'10}_1 \cos(\theta) \ , \nonumber \\ 
{\cal H}_5 & = & 0 \ , \ \ \ \ 
{\cal H}_6  =  C^{'01}_0  \ , \ \ \ \
{\cal H}_7  = 0 \ , \ \ \ \ \ \ \ \   
{\cal H}_8  = 0 \ , 
\label{SSTRUC1_nam_pw}
\end{eqnarray}
where the quantities $A^{'S'S}_{L'}$, $B^{'S'S}_{L'}$, etc, are given in Appendix C.

An interesting feature in Eq.(\ref{SSTRUC1_pam_pw}) 
is that the amplitudes with spin-triplet initial states depend linearly on 
$(p'/\Lambda)$ (i.e., the final state is in a $P$ wave) if the parity of the baryon 
$B$ is positive, whereas if the parity of  $B$ is negative 
(Eq.(\ref{SSTRUC1_nam_pw})), they do not depend on $(p'/\Lambda)$ (i.e., the final 
state is an $S$ state). This energy dependence is 
interchanged for amplitudes with spin-singlet initial states. The feature just 
mentioned is a direct consequence of the Pauli principle and total parity conservation, 
\begin{equation}
(-)^{S+L'+T}= \mp 1 \ , 
\label{Pauli}
\end{equation}
where the ($-$) and ($+$) signs refer to the positive and negative parity baryon $B$,
respectively. $T$ denotes the total isospin ($T=1$ for $pp$). In fact, as pointed 
out in Ref.\cite{HHNM}, it follows immediately from Eq.(\ref{Pauli}) that the energy 
dependence of a partial-wave amplitude with spin-triplet initial state ($S=1$) is given 
by an odd power of $(p'/\Lambda)$ if the parity of $B$ is positive since in this 
case the final state must be in an odd orbital angular momentum $L'$, whereas it is 
given by an even power of $(p'/\Lambda)$ if the parity is negative since $L'$ is even 
in this case. For completeness, we note that for the $pn \to B' B$ 
reaction, this energy dependence is interchanged \cite{HHNM}.

With the coefficients ${\cal D}_i$ and ${\cal H}_i$ given by 
Eqs.(\ref{SSTRUC1_pam_pw},\ref{SSTRUC1_nam_pw}), it is straightforward to 
calculate any observable of interest in the $pp \to B'B$ reaction using 
Eqs.(\ref{ampl_obsp},\ref{ampl1_obsp},\ref{ampl_obsn},\ref{ampl1_obsn}) and the
method given in Appendix D. Alternatively, one can calculate the observables using 
the matrix elements given by Eq.(\ref{PWE}), in terms of which, some of the observables 
are also given in Appendix D. 
In particular, the cross section with the spin-triplet initial state $^3\sigma_\Sigma$,  
as defined by Eq.(\ref{xsc30}) in Appendix D, can be expressed as
\begin{equation}
 \left(\frac{p'}{\Lambda}\right)^{-1} \frac{d( ^3\sigma^+_\Sigma)}{d\Omega} = 
\left[\beta_0 + \beta_1\cos^2(\theta)\right]
 \left(\frac{p'}{\Lambda}\right)^2 \ ,
\label{xsc3+}
\end{equation}
for a positive parity baryon $B$ and,    
\begin{equation}
\left(\frac{p'}{\Lambda}\right)^{-1} \frac{d (^3\sigma^-_\Sigma)}{d\Omega} = \beta_0^\prime \ , 
\label{xsc3-}
\end{equation}
for a negative parity baryon $B$. Note that we have divided out the cross section 
by a factor of $(p'/\Lambda)$ due to the final state phase space
which introduces an extra $p'$ dependence into the cross section. In the above 
equations
\begin{eqnarray}
\beta_0 & \equiv & 2|B^1_1|^2 + \frac{1}{2}\left(|C^1_1|^2 + |E^1_1|^2\right)  
- |U^1_1|^2 \ , \nonumber \\ 
\beta_1 & \equiv & - \beta_0 + \frac{3}{4}|A^1_1|^2
+ \frac{2}{3}|D^1_1 + E^1_1|^2 - 2|W^1_1|^2 \ , \nonumber \\
\beta_0^\prime & \equiv & 2|A'^{11}_0|^2 + |C'^{01}_0|^2  \ ,
\label{eadep}
\end{eqnarray}
where $U^1_1$ and $W^1_1$ are defined in Appendix C.

Eqs.(\ref{xsc3+},\ref{xsc3-}) show that, apart from the $p'$ dependence due to 
phase space, the spin-triplet cross section scales 
quadratically in $p'$ (or equivalently, linearly in excess energy $Q$ since 
$\sqrt{Q} \propto p'$), if the parity of $B$ is positive, whereas it 
is constant if the parity of $B$ is negative. This is the principal result 
of Ref.\cite{HHNM}, where the reaction $NN \to Y\Theta^+$ has been investigated. 
Furthermore, the spin-triplet cross section exhibits a 
$\cos^2(\theta)$ angular dependence in the case of a positive parity $B$, 
whereas it is isotropic in the case of a negative parity $B$.  Therefore,
an observation of a strong $cos^2(\theta)$ dependence in the measured angular
distribution near threshold would imply a positive parity baryon $B$. 
An isotropic angular distribution, on the other hand, would be inconclusive about
the parity of $B$. 

The spin-triplet cross section can be related to the spin correlation 
coefficients $A_{ii}$ as (see Appendix D)
\begin{equation}
\frac{d(^3\sigma_\Sigma)}{d\Omega} = \frac{1}{4} \frac{d\sigma}{d\Omega}
(2 + A_{xx} + A_{yy}) \ ,
\label{xsc3}
\end{equation}
with $d\sigma/d\Omega$ denoting the unpolarized cross section and,
$A_{ii}$, the spin correlation coefficient with the spin orientations along the 
$i$-axis. 
Therefore, in general, $^3\sigma_\Sigma$ can be extracted experimentally by measuring the
spin correlation coefficients $A_{xx}$ and $A_{yy}$ in conjunction with the unpolarized
cross section. Furthermore, it is immediate from Eq.(\ref{Cij_SME}) that, at 
threshold, as pointed out by Rekalo and Tomasi-Gustafsson in Ref.\cite{Hadronic} in 
connection to the reaction $pp \to \Sigma^+\Theta^+$, $A_{xx}=A_{yy}=-1$ for a positive 
parity baryon $B$ while $A_{xx}=A_{yy} \geq 0$ for a negative parity $B$.

\section{The reaction $N + N \to M + B' + N$}
We now focus on the process $N + N \to M + B' + N$, where $M$ stands for a spinless 
(scalar or pseudoscalar) meson and, $B'$ for a spin-1/2 and positive-parity baryon. 
We start by making a partial-wave expansion of the corresponding reaction amplitude
\begin{eqnarray}
<S'M_{S'}|\hat M(\vec q, \vec p\ '; \vec p)|SM_S>  = 
& \sum & i^{L-L'-l} (lm_lJ'M_{J'}|JM_J)(S'M_{S'}L'M_{L'}|J'M_{J'})\nonumber \\
& \times & (SM_SLM_L|JM_J) M^{S'J'SJ}_{lL'L}(q, p'; p) \nonumber \\ 
& \times & Y_{lm_l}(\hat q) Y_{L'M_{L'}}(\hat p') Y^*_{LM_L}(\hat p) \ , 
\label{PWE_MNN}
\end{eqnarray}
where $S, L, J$ stand for the total spin, total orbital angular momentum and the total 
angular momentum, respectively, of the initial $NN$ state. $M_S$, $M_L$ and $M_J$ denote 
the corresponding projection quantum numbers. The primed quantities stand for the 
corresponding quantum numbers of the final $B'N$ state. $l$ and $m_l$ denote the orbital 
angular momentum of the emitted meson and its projection, respectively, relative to the
center-of-mass of the final two-baryon system. The summation runs over all quantum 
numbers not specified in the l.h.s. of Eq.(\ref{PWE_MNN}). $\vec p$ and 
$\vec p\ '$ denote the relative momenta of the two baryons in the initial and final 
states, respectively. $\vec q$ denotes the momentum of the emitted meson with respect to 
the center-of-mass of the two baryons in the final state. We note that, in Eq.(\ref{PWE_MNN}), 
apart from the restrictions on the quantum numbers encoded in the geometrical factors, 
total parity conservation imposes that $(-)^{l+L+L'} = - 1$ in the case of pseudoscalar 
meson production and $(-)^{l+L+L'} = + 1$ in the case of scalar meson production.

Eq.(\ref{PWE_MNN}) can be inverted to solve for the partial-wave matrix element
$M^{S'J'SJ}_{lL'L}(q, p'; p)$. We have
\begin{eqnarray}
M^{S'J'SJ}_{lL'L}(q, p'; p) & = &
\sum i^{L'+l-L}  (lm_lJ'M_{J'}|JM_J)(S'M_{S'}L'M_{L'}|J'M_{J'})(SM_SL0|JM_J)\nonumber \\
& \times & \frac{8\pi^2}{2J+1} \sqrt{\frac{2L+1}{4\pi}}
\int d\Omega_{p'} Y_{L'M_{L'}}^*(\hat p') \nonumber \\
& \times & 
\int_{-1}^{+1} d(\cos(\theta_q)) Y_{lm_l}^*(\theta_q, 0)
<S'M_{S'}|\hat M(\vec q, \vec p\ '; \vec p)|SM_S> \ ,
\label{pwme_MNN}
\end{eqnarray}
where, without loss of generality, the $z$-axis is chosen along $\vec p$ and  
$\vec q$ in the $xz$-plane; $\cos(\theta_q) = \hat q \cdot \hat p$. The summation 
is over all quantum numbers not specified in the l.h.s. of the equation.  

The most general spin structure of the transition operator can be extracted from 
Eq.(\ref{PWE_MNN}) as
\begin{equation}
\hat M(\vec q, \vec p\ '; \vec p) = \sum_{S'SM_{S}M_{S'}} 
|S'M_{S'}> <S'M_{S'}|\hat M(\vec q, \vec p\ '; \vec p)|SM_S> <SM_S| \ .
\label{SSTRUC0_MNN}
\end{equation}
Inserting Eq.(\ref{PWE_MNN}) into Eq.(\ref{SSTRUC0_MNN}) and re-coupling gives
\begin{eqnarray}
\hat M(\vec q, \vec p\ '; \vec p) & = & \sum i^{L-L'-l} (-)^{L-J+J'+l+S'+S} [J'][J]^2  
M^{S'J'SJ}_{lL'L}(q, p'; p) \nonumber \\  & \times  &
\sum_{\alpha\beta}
  \begin{Bmatrix}
  L' & J' & S' \\   J & \beta & l
  \end{Bmatrix}
  \begin{Bmatrix}
  S' & \beta & J \\   L & S & \alpha
  \end{Bmatrix}
[A_{S'} \otimes B_S]^\alpha \cdot [X_{(lL')\beta\ L}]^\alpha \ ,
\label{SSTRUC1_MNN}
\end{eqnarray}
where $[X_{(lL')\beta\ L}]^\alpha$ is defined as
\begin{equation}
[X_{(lL')\beta\ L}]^\alpha \equiv
\left[ [Y_{l}(\hat q) \otimes Y_{L'}(\hat p')]^\beta \otimes Y_L(\hat p) \right]^\alpha  \ ,
\label{aux1_MNN}
\end{equation}
and contains all the information on the angular dependence of the transition operator. 

We now expand $[A_{S'} \otimes B_S]^\alpha$, for each tensor rank $\alpha$, 
in terms of the complete set of available spin operators in the problem.
Since the meson $M$ produced in the final state is a spinless meson, the 
expansion of $[A_{S'} \otimes B_S]^\alpha$ is exactly the same as that for the 
$N + N \to B' + B$ reaction discussed in section III and is given by Eq.(\ref{SOPER}).

Inserting Eq.(\ref{SOPER}) into Eq.(\ref{SSTRUC1_MNN}) yields
\begin{eqnarray}
\hat M(\vec q, \vec p\ '; \vec p) & = & {\cal R}_0 P_{S=0} +  {\cal R}_1 P_{S=1} 
+ \vec {\cal R}_2 \cdot (\vec\sigma_1 + \vec\sigma_2)
+ \vec {\cal R}_3 \cdot (\vec\sigma_1 - \vec\sigma_2) P_{S=0}
+ \vec {\cal R}_4 \cdot (\vec\sigma_1 - \vec\sigma_2) P_{S=1} \nonumber \\
& + & 
{\cal R}^2_5 \cdot [\vec\sigma_1 \otimes \vec\sigma_2]^2 \ . 
\label{SSTRUC2_MNN}
\end{eqnarray}
The above result is the most general spin structure of the spinless meson production
operator in $NN$ collisions consistent with symmetry principles.
The first two terms are the central spin singlet and triplet interactions, 
respectively. The third term is a tensor of rank 1 and corresponds to the usual spin-orbit 
interaction. The fourth and fifth terms are also tensors of rank 1 but they describe the spin 
singlet$\to$triplet and triplet$\to$singlet transitions, respectively. The last term corresponds 
to the tensor interaction of rank 2. Apart from the fourth and fifth terms, all the other terms 
conserve the total spin in the transition. The coefficients multiplying the spin operators in 
the above equation contain the dynamics of the reaction process and are given by
\begin{eqnarray}
{\cal R}_0 & = & \sum i^{L-L'-l} (-)^{L} [L] M^{0L'0L}_{lL'L}(q, p'; p) [X_{(lL')L\ L}]^0 \ , \nonumber \\
{\cal R}_1 & = & \frac{1}{3} \sum i^{L-L'-l} (-)^{J'} [J'] [J]^2 
  \begin{Bmatrix}
  L' & J' & 1 \\   J & L & l
  \end{Bmatrix}
M^{1J'1J}_{lL'L}(q, p'; p) [X_{(lL')L\ L}]^0 \ , \nonumber \\
\vec {\cal R}_2 & = & 
\frac{1}{2\sqrt{2}} \sum i^{L-L'-l} (-)^{L-J+J'+1} [J'][J]^2 M^{1J'1J}_{lL'L}(q, p'; p) \nonumber \\
& \times & \sum_{\beta} [\beta] 
  \begin{Bmatrix}
  L' & J' & 1 \\   J & \beta & l
  \end{Bmatrix}
  \begin{Bmatrix}
  1 & \beta & J \\   L & 1 & 1
   \end{Bmatrix} 
[X_{(lL')\beta\ L}]^1 \ , \nonumber \\
\vec {\cal R}_3 & = & 
\frac{1}{2\sqrt{3}} \sum i^{L-L'-l} (-)^{L+J'} [L][J'] M^{1J'0L}_{lL'L}(q, p'; p)
\sum_{\beta} (-)^\beta [\beta] 
  \begin{Bmatrix}
  L' & J' & 1 \\   L & \beta & l
  \end{Bmatrix} 
[X_{(lL')\beta\ L}]^1 \ , \nonumber \\
\vec {\cal R}_4 & = & 
\frac{1}{2\sqrt{3}} \sum i^{L-L'-l} (-)^{L'-J+1} [J]
M^{0L'1J}_{lL'L}(q, p'; p) [X_{(lL')J\ L}]^1 \ , \nonumber \\
{\cal R}^2_5 & = & 
\frac{1}{2} \sum i^{L-L'-l} (-)^{L-J+J'+1} [J'][J]^2 M^{1J'1J}_{lL'L}(q, p'; p) \nonumber \\
& \times & \sum_{\beta} [\beta] 
  \begin{Bmatrix}
  L' & J' & 1 \\   J & \beta & l
  \end{Bmatrix}
  \begin{Bmatrix}
  1 & \beta & J \\   L & 1 & 2
  \end{Bmatrix} 
[X_{(lL')\beta\ L}]^2  \ . 
\label{COEFF_MNN}
\end{eqnarray}
Note that the summations over the quantum numbers in the above equation are restricted by
total parity conservation and the Pauli principle:
\begin{eqnarray}
(-)^{l+L+L'} & = & \pm 1 \ , \nonumber \\
(-)^{L+S+T}  & = & -1 \ ,
\label{SYM1_MNN}
\end{eqnarray}
where the $\pm$ signs refer to scalar $(+)$ and pseudoscalar $(-)$ mesons, respectively;
$T$ denotes the total isospin of the two interacting nucleons in the initial state.
If the baryon $B$ in the final state is a nucleon, the summations in Eq.(\ref{COEFF_MNN}) 
are further restricted by the Pauli principle
\begin{equation}
(-)^{L'+S'+T'} = -1 \ ,
\label{SYM2_MNN}
\end{equation}
where $T'$ denotes the total isospin of the two interacting nucleons in the final state.

The quantity $[X_{(lL')\beta\ L}]^\alpha$ in the above equation can be most easily evaluated 
by choosing the $z$-axis along the relative momentum $\vec p$ of 
the two nucleons in the initial state. We then have
\begin{eqnarray}
{[X_{(lL')\beta\ L}]}^\alpha_{M_\alpha} &=& \frac{[LL']}{\sqrt{4\pi}} (\beta M_\alpha L0|\alpha M_\alpha)
[Y_{l}(\hat q) \otimes Y_{L'}(\hat p')]^\beta \nonumber \\
 &=& \frac{[LL']}{\sqrt{4\pi}} (\beta M_\alpha L0|\alpha M_\alpha)
\sum_{m_l,M_{L'}} (lm_lL'M_{L'}|\beta M_\alpha) Y_{lm_l}(\hat q) Y_{L'M_{L'}}(\hat p\ ') \ .
\label{ang_n_MNN}
\end{eqnarray}
The evaluation of $[X_{(lL')^\beta\ L}]^\alpha$ can be further simplified if we choose the  
relative momentum $\vec p\ '$ of the two nucleons in the final state in the $xz$-plane, in
which case, 
$Y_{L'M_{L'}}(\hat p\ ') = (-)^{M_{L'}} \sqrt{(L'- M_{L'})!/ (L'+ M_{L'})!} [(2L'+1)/4\pi]^{1/2}
P^{M_{L'}}_{L'}(\hat p\ ' \cdot \hat p)$.

What we have done so far is completely general and applies to any spinless meson 
production in $NN$ collisions. In the following, we treat the production of neutral and charged 
spinless mesons separately in the $NN\to MNN$ reaction.

\subsection{Neutral meson production amplitude in $N+N\to M+N+N$.}

Since the only difference between the scalar and pseudoscalar meson production amplitude is 
that the sum $l+L+L'$ be even (scalar meson) or odd (pseudoscalar meson) as expressed in 
Eq.(\ref{SYM1_MNN}), we restrict the following considerations to {\it pseudoscalar} meson production.
The scalar meson production amplitude is easily obtained from the pseudoscalar meson production 
amplitude by changing the restriction on the orbital angular momenta from $(-)^{l+L+L'}=-1$
to $(-)^{l+L+L'}=+1$.

In the case of neutral meson production, the available isospin operators in the problem are 
the usual isospin operators $\vec\tau_1$ and $\vec\tau_2$ together with the identity operator
acting on the nucleon sector. We need to construct a scalar from these available operators. 
Since the total isospin of the two interacting nucleons is conserved in the production 
of a neutral meson, the isospin structure of the reaction amplitude is most simply expressed in
terms of the isospin projection operator $P_T$ for the iso-singlet and iso-triplet transitions 
as $T=0$ and $1$, respectively. Explicitly, $P_{T=0} = (1 - \vec\tau_1\cdot\vec\tau_2) / 4$ 
and $P_{T=1} = (3 + \vec\tau_1\cdot\vec\tau_2) / 4$. 

We now note that, since the total isospin is conserved in the neutral meson production process, 
total parity conservation together with Pauli principle (Eqs.(\ref{SYM1_MNN},\ref{SYM2_MNN})) 
demands that $(-)^{l+S+S'}=-1$ in the case of pseudoscalar meson production. This implies that the 
coefficients $\vec {\cal R}_3$ and $\vec {\cal R}_4$ in Eq.(\ref{COEFF_MNN}) involve the summation over even $l$ 
only while all other coefficients in Eq.(\ref{COEFF_MNN}) involve summation over odd $l$ only. In 
what follows we will use the superscript $(e)$ or $(o)$ in the coefficients in Eq.(\ref{COEFF_MNN}) 
to indicate the restricted summation to even or odd $l$, respectively.

With the above considerations, the (pseudoscalar) neutral meson production operator for a 
transition with a given isospin $T$ is given by 
\begin{eqnarray}
\hat M_T(\vec q, \vec p\ '; \vec p) & = & \Big\{ {\cal R}^{(o)}_0 P_{S=0} +  {\cal R}^{(o)}_1 P_{S=1} 
+ \vec {\cal R}^{(o)}_2 \cdot (\vec\sigma_1 + \vec\sigma_2)
+ \vec {\cal R}^{(e)}_3 \cdot (\vec\sigma_1 - \vec\sigma_2) P_{S=0} \nonumber \\
& + & \vec {\cal R}^{(e)}_4 \cdot (\vec\sigma_1 - \vec\sigma_2) P_{S=1}
+ {\cal R}^{2(o)}_5 \cdot [\vec\sigma_1 \otimes \vec\sigma_2]^2 \Big\} P_T \ . 
\label{SSTRUC1_nT_MNN}
\end{eqnarray}
The above result is the most general spin-isospin structure of the neutral pseudoscalar meson 
production operator in $NN$ collisions consistent with symmetry principles. 

In the total isospin basis, the $pp$ and $pn$ states are expressed as
\begin{eqnarray}
|pp> & = & |T=1, M_T=1> \ , \nonumber \\
|pn> & = & \frac{1}{\sqrt{2}} [|T=1, M_T=0> + |T=0, M_T=0>] \ ,
\label{NN_T_MNN}
\end{eqnarray}
so that, for $pp$ collisions, we have
\begin{equation}
\hat M_{ppM}(\vec q, \vec p\ '; \vec p) = \hat M_{T=1}(\vec q, \vec p\ '; \vec p) \ ,
\label{SSTRUC_pp_MNN}
\end{equation}
while for $pn$ collisions,  
\begin{equation}
\hat M_{pnM}(\vec q, \vec p\ '; \vec p) = \frac{1}{2} \sum_{T=0,1} \hat M_T(\vec q, \vec p\ '; \vec p)\ .
\label{SSTRUC_pn_MNN}
\end{equation}

The transition operator for neutral {\it scalar} meson production is also given by Eq.(\ref{SSTRUC1_nT_MNN}), 
except that the superscripts $(e)$ and $(o)$ in the quantities appearing in Eq.(\ref{SSTRUC1_nT_MNN}) 
should be interchanged. 

\subsection{Charged meson production amplitude in $N+N\to M+N+N$.}

For charged meson production, the available nucleon isospin operators are the same
as those for neutral meson production as discussed in the previous subsection. In addition,
there is the isospin creation operator $\hat\pi^\dagger_m$ in the meson sector [$|1m> = \hat\pi^\dagger_m|0>$], 
where the subscript $m$ stands for the charge of the produced meson ($m=+1,0,-1$). Again, we 
form all possible scalars from these operators with $\hat\pi_m$ appearing once 
in each term. The isospin structure of the transition operator is then of the form 
$\hat\pi \cdot \hat O$, where $\hat O$ stands for an isospin operator of rank 1 
in the two nucleon isospin sector. Three independent structures are possible for $\hat O$:
\begin{eqnarray}      
T=T'=1 :  \ \ \ \ \ \ \ \
\hat O & = &  \left( \vec\tau_1 + \vec\tau_2 \right)  
\ , \nonumber \\
T=0, T'=1 :  \ \ \ \ \ \ \ \ 
\hat O  & = & \left( \vec\tau_1 - \vec\tau_2 \right) P_{T=0} \ ,
\nonumber \\
T=1, T'=0 : \ \ \ \ \ \ \  \
\hat O  & = & \left( \vec\tau_1 - \vec\tau_2 \right) P_{T=1} \ .
\label{TOPER_MNN}
\end{eqnarray}

In the isospin singlet$\to$triplet and triplet$\to$singlet transitions, total parity 
conservation and the Pauli principle (Eqs.(\ref{SYM1_MNN},\ref{SYM2_MNN})) demand that 
$(-)^{l+S+S'}=+1$ in the case of pseudoscalar meson production.
This condition is just opposite to that for either the isospin singlet-singlet or 
triplet-triplet transitions as discussed in the previous subsection. Taking into account 
this restriction together with Eq.(\ref{TOPER_MNN}), we have for a given transition $T \to T'$
\begin{eqnarray}
\hat M_{T'T}(\vec q, \vec p\ '; \vec p) & = & \Big\{ {\cal R}^{(o)}_0 P_{S=0} +  {\cal R}^{(o)}_1 P_{S=1} 
+ \vec {\cal R}^{(o)}_2 \cdot (\vec\sigma_1 + \vec\sigma_2)
+ \vec {\cal R}^{(e)}_3 \cdot (\vec\sigma_1 - \vec\sigma_2) P_{S=0} \nonumber \\
& + & \vec {\cal R}^{(e)}_4 \cdot (\vec\sigma_1 - \vec\sigma_2) P_{S=1}
+ {\cal R}^{2(o)}_5 \cdot [\vec\sigma_1 \otimes \vec\sigma_2]^2 \Big\}
\hat \pi \cdot (\vec\tau_1 + \vec\tau_2)  \nonumber \\
& + & 
\Big\{ {\cal R}^{(e)}_0 P_{S=0} +  {\cal R}^{(e)}_1 P_{S=1} 
+ \vec {\cal R}^{(e)}_2 \cdot (\vec\sigma_1 + \vec\sigma_2)
+ \vec {\cal R}^{(o)}_3 \cdot (\vec\sigma_1 - \vec\sigma_2) P_{S=0} \nonumber \\
& + & \vec {\cal R}^{(o)}_4 \cdot (\vec\sigma_1 - \vec\sigma_2) P_{S=1}
+ {\cal R}^{2(e)}_5 \cdot [\vec\sigma_1 \otimes \vec\sigma_2]^2 \Big\}
\hat \pi \cdot (\vec\tau_1 - \vec\tau_2) P_T \ .
\label{SSTRUC1_cT_MNN}
\end{eqnarray}
Note that the coefficients of the spin operators in the two-nucleon isospin non-flip
term have different restrictions on the angular momentum of the emitted meson relative 
to the corresponding coefficients in the two-nucleon isospin flip
term. The above result is the most general spin-isospin structure of the charged 
pseudoscalar meson production operator in $NN$ collisions consistent with 
symmetry principles. The transition operator for a specific reaction can be trivially obtained
from Eq.(\ref{SSTRUC1_cT_MNN}) using Eq.(\ref{NN_T_MNN}).

The transition operator for charged scalar meson production is also given by Eq.(\ref{SSTRUC1_cT_MNN}), 
except that the superscripts $(e)$ and $(o)$ in the quantities appearing in 
Eq.(\ref{SSTRUC1_cT_MNN}) should be interchanged.

\section{Near-threshold $pp \to ppM$ reaction.}

In any particle production reaction, the near-threshold energy region is of particular 
interest due to the limited number of relevant partial-wave amplitudes.
In this section, as an example, we consider the $pp \to ppM$ reaction
near threshold, where the produced pseudoscalar meson $M$ is primarily in an $s$-wave state. 
Then, considering only $l=0$, the transition operators of the previous sections take particularly 
simple forms. Indeed, Eq.(\ref{SSTRUC_pp_MNN}) becomes 
\begin{equation}
\hat M_{ppM}(\vec q, \vec p\ '; \vec p)  =  
\vec {\cal R}^{(e)}_3 \cdot (\vec\sigma_1 - \vec\sigma_2) P_{S=0} + 
\vec {\cal R}^{(e)}_4 \cdot (\vec\sigma_1 - \vec\sigma_2) P_{S=1} \ ,
\label{SSTRUC1_na_MNN}
\end{equation}
where the isopin matrix element has been already taken. From Eq.(\ref{COEFF_MNN}), the coefficients 
$\vec {\cal R}^{(e)}_3$ and $\vec {\cal R}^{(e)}_4$ reduce to
\begin{eqnarray}
\vec {\cal R}^{(e)}_3  & = & \frac{1}{2\sqrt{3}} \sum i^{L-L'} (-)^{L'} [L] M^{1L0L}_{0L'L}(q, p'; p)
[X_{(0L')L'\ L}]^1 \ , \nonumber \\
\vec {\cal R}^{(e)}_4  & = & - \frac{1}{2\sqrt{3}} \sum i^{L-L'} [L']
M^{0L'1L'}_{0L'L}(q, p'; p) [X_{(0L')L'\ L}]^1 \ ,
\label{SSTRUC1_na1_MNN}
\end{eqnarray}
with
\begin{equation}
[X_{(0L')L'\ L}]^1 = \frac{[LL']}{(4\pi)^{\frac{3}{2}}}
\left[ \sqrt{\frac{2}{L'(L'+1)}} (L'1L0|11) P_{L'}^1(\hat p \cdot \hat p') \hat n_1
+ (L'0L0|10)P_{L'}(\hat p \cdot \hat p') \hat p \right] \ .
\label{ang_na_MNN}
\end{equation}
In the above equation, $\hat n_1 \equiv [(\vec p \times \vec p\ ') \times \vec p ]/ 
|\vec p \times \vec p\ '|$. 

Restricting the final two proton states to $S$ and $P$ waves, Eq.(\ref{SSTRUC1_na_MNN}) 
reduces further to,  
\begin{equation}
\hat M_{ppM}(\vec q, \vec p\ '; \vec p)  =  
i \Big\{ \left[ \beta_2\ \hat p' -  \beta_3\ (\hat p' - 3\hat p \cdot \hat p' \hat p) \right] 
\cdot (\vec\sigma_1 - \vec\sigma_2) P_{S=0}  
- \beta_1\ \hat p  \cdot (\vec\sigma_1 - \vec\sigma_2) P_{S=1}
\Big\}  \ , 
\label{SSTRUC1_naSP_MNN}
\end{equation}
where 
\begin{eqnarray}
\beta_1 & \equiv & \frac{1}{2(4\pi)^{3/2}} M^{1101}_{001}(q, p'; p) \ , \nonumber \\
\beta_2 & \equiv & \frac{1}{2(4\pi)^{3/2}} M^{0111}_{010}(q, p'; p) \ , \nonumber \\
\beta_3 & \equiv & \frac{1}{2(4\pi)^{3/2}}\sqrt{\frac{5}{2}} M^{0111}_{012}(q, p'; p) \ , 
\label{aaa_MNN}
\end{eqnarray}
are proportional to the partial-wave matrix elements corresponding to 
$^3P_0 \to\,^1S_0 s$, $^1S_0 \to\,^3P_0 s$, and $^1D_2 \to\,^3P_2 s$ transitions, 
respectively \cite{notation}. Eq.(\ref{SSTRUC1_naSP_MNN}) agrees with the structure 
derived in Ref.\cite{Nak1} based directly on symmetry considerations. We also 
note that the higher partial-wave contributions in $pp \to pp\eta$ reported 
in Ref.\cite{Nak1} have been calculated based on the present method.

The transition operator given by Eqs.(\ref{SSTRUC1_nT_MNN},\ref{SSTRUC_pp_MNN}) can be 
re-expressed in the form
\begin{equation}
\hat M_{ppM} = \sum_{\lambda=1}^6 \sum_{n,n'=0}^3 
M^\lambda_{nn'} \sigma_n(1) \sigma_{n'}(2) \ ,
\label{ampl_MNN}
\end{equation}
where
\begin{eqnarray}
M^1_{nn'} & = & \frac{1}{4} (3{\cal R}^{(o)}_1 + {\cal R}^{(o)}_0) \delta_{n,0} \delta_{n',0} \ , \nonumber \\
M^2_{nn'} & = & \left\{ \frac{1}{4} ({\cal R}^{(o)}_1 - {\cal R}^{(o)}_0) + \tilde {\cal R}^{2(o)}_{a5} \right\}
(1-\delta_{n,0}) \delta_{n,n'} \ , \nonumber \\
M^3_{nn'} & = & [\vec {\cal R}^{(o)}_2 + \frac{1}{2}(\vec {\cal R}^{(e)}_3 + \vec {\cal R}^{(e)}_4)]_n 
(1-\delta_{n,0}) \delta_{n',0} \ , \nonumber \\
M^4_{nn'} & = & [\vec {\cal R}^{(o)}_2 - \frac{1}{2}(\vec {\cal R}^{(e)}_3 + \vec {\cal R}^{(e)}_4)]_{n'} 
(1-\delta_{n',0}) \delta_{n,0} \ , \nonumber \\
M^5_{nn'} & = & i \frac{1}{2} \sum_k \varepsilon_{nn'k} (\vec {\cal R}^{(e)}_3 - \vec {\cal R}^{(e)}_4)_k 
(1-\delta_{n,0}) (1-\delta_{n',0}) \ , \nonumber \\
M^6_{nn'} & = &  \tilde {\cal R}^{2(o)}_{b5} (1-\delta_{n,0}) (1-\delta_{n',0}) \ ,
\label{ampl1_MNN}
\end{eqnarray}
with 
\begin{eqnarray}
\tilde {\cal R}^{2(o)}_{a5} & \equiv & \sum_{m_1m_2m} (-)^m {\cal R}^{2(o)}_{5m} (1m_1 1m_2 | 2-m)
\nonumber \\ & \times &                   
\left\{ \frac{m_1m_2}{2}(\delta_{n,1} - m_1m_2\delta_{n,2}) +
(1-|m_1|)(1-|m_2|)\delta_{n,3} \right\} \ , \nonumber \\
\tilde {\cal R}^{2(o)}_{b5} & \equiv & \sum_{m_1m_2m} (-)^m {\cal R}^{2(o)}_{5m} (1m_1 1m_2 | 2-m)                    
\nonumber \\ & \times &
\Big\{ i \frac{m_1m_2}{2}(m_2\delta_{n,1}\delta_{n',2} + m_1\delta_{n,2}\delta_{n',1})
- \frac{m_1}{\sqrt{2}} (1-|m_2|) (\delta_{n,1} + i m_1 \delta_{n,2})\delta_{n',3} \nonumber \\ & - &
  \frac{m_2}{\sqrt{2}} (1-|m_1|) \delta_{n,3}(\delta_{n',1} + i m_2 \delta_{n',2}) \Big\} \ .
\label{ampl2_MNN}
\end{eqnarray}

Any observable of interest can then be calculated either from Eqs.(\ref{ampl_MNN}) 
or from Eq.(\ref{PWE_MNN}) as given in Appendix D.

In near-threshold kinematics with only $s$-wave mesons, we see from  
Eqs.(\ref{SSTRUC1_na_MNN},\ref{ampl1_MNN}) that the non-vanishing coefficients are 
$M^{\lambda=3,4,5}_{nn'}$. If, in addition, the final two nucleon states are restricted 
to $S$ and $P$ waves, we have 
\begin{eqnarray}
M^3_{nn'} & = & i\frac{1}{2} \Big\{ (\beta_2 - \beta_3) \hat p'_n +
[3\beta_3(\hat p \cdot \hat p') - \beta_1]\hat p_n \Big\} (1-\delta_{n,0}) \delta_{n',0} \ , \nonumber \\
M^4_{nn'}& = &  - i \frac{1}{2}\Big\{ (\beta_2 - \beta_3) \hat p'_{n'} + 
[3\beta_3(\hat p \cdot \hat p') - \beta_1]\hat p_{n'} \Big\} \delta_{n,0} (1-\delta_{n',0}) \ , \nonumber \\
M^5_{nn'}& = & - \frac{1}{2}
\sum_k \varepsilon_{nn'k} \Big\{ (\beta_2 - \beta_3) \hat p'_k + 
[3\beta_3(\hat p \cdot \hat p') + \beta_1]\hat p_k \Big\} (1-\delta_{n,0}) (1-\delta_{n',0}) \ , 
\label{ampl1_thSP_MNN}
\end{eqnarray}
where $\beta_i$'s are defined in Eq.(\ref{aaa_MNN}).

Inserting the above results into Eqs.(\ref{xsc},\ref{Ai},\ref{Cij}) (or alternatively using 
Eq.(\ref{PWE_MNN}) and Eqs.(\ref{xsc_SME},\ref{Ai_SME},\ref{Cij_SME})) yields
\begin{eqnarray}
\frac{d\sigma}{d\Omega} & = & |\beta_1|^2 + |\beta_2|^2 + (3\cos^2\theta + 1)|\beta_3|^2
+ 2(3\cos^2\theta - 1)\Re[\beta_2\beta_3^*] \ , \nonumber \\
\frac{d\sigma}{d\Omega} A_i & = & 0 \ , \nonumber \\
\frac{d\sigma}{d\Omega} A_{xx} & = & \frac{d\sigma}{d\Omega} A_{yy} \nonumber \\ 
& = & |\beta_1|^2 - |\beta_2|^2 - (3\cos^2\theta + 1)|\beta_3|^2
- 2(3\cos^2\theta - 1)\Re[\beta_2\beta_3^*] \ ,
\label{OBS_th_MNN}
\end{eqnarray}
which are equivalent to the results derived directly from Eq.({\ref{SSTRUC1_naSP_MNN}) 
in Ref.\cite{Nak1}. From the above results, the final state ($NN$) $S$-wave 
contribution can be isolated via the combination:
\begin{equation}
\frac{1}{2}\frac{d\sigma}{d\Omega}(1+A_{xx}) = \frac{d(^3\sigma)}{d\Omega}
= |\beta_1|^2 \ ,
\end{equation}
which is nothing other than a consequence of the Pauli principle and parity
conservation as discussed
in section IV.

\section{Summary.}

Based on the partial-wave expansion of the reaction amplitude, we have derived the 
most general spin structure of the transition operator for the reactions 
$\gamma + N \to M + B$ and $N + N \to B' + B$. Also, we have derived the most general 
spin structure of the spinless meson production operator for the $N + N \to M + B' + N$ 
reaction. The present method used to extract the spin structure of the transition operator 
is quite general and, in principle, can be applied to any reaction process in a systematic 
way. The advantage of this method is that it relates the coefficients multiplying each 
spin operator directly to the partial-wave matrix elements to any desired order of the 
corresponding expansion.

\vspace{2em}
\noindent
{\bf Acknowledgment:}\\
\noindent
This work is supported by Forschungszentrum-J\"{u}lich, 
contract No. 41445282 (COSY-058).

\section{Appendix A}
In this appendix we will determine the coefficients $a_{L'L}$, $b_{L'L}$ and $c_{L'L}$
in Eq.(\ref{aux11}) as well as $a'_{L'L}$ and $b'_{L'L}$ in Eq.(\ref{aux11n}).

Taking the scalar product of the last equality in Eq.(\ref{aux11}) with 
$[\hat q \otimes \hat q]^2$, $[\hat k \otimes \hat k]^2$ and $[\hat k \otimes \hat q]^2$, 
respectively, we have
\begin{eqnarray}
3 u & = & 2 a_{L'L} + (3\cos^2\theta-1) b_{L'L} + 2\cos\theta c_{L'L}  \nonumber \\
3 v & = & (3\cos^2\theta-1) a_{L'L} + 2 b_{L'L} + 2\cos\theta c_{L'L}  \nonumber \\
3 w & = & 2\cos\theta a_{L'L} + 2\cos\theta b_{L'L} + \frac{1}{2}(3+\cos^2\theta) c_{L'L} \ ,
\label{a1}
\end{eqnarray}
where $\cos\theta \equiv \hat k \cdot \hat q$ and
\begin{eqnarray}
u &\equiv & [Y_L(\hat k) \otimes Y_{L'}(\hat q)]^2 \cdot [\hat q \otimes \hat q]^2 \nonumber \\
v &\equiv & [Y_L(\hat k) \otimes Y_{L'}(\hat q)]^2 \cdot [\hat k \otimes \hat k]^2 \nonumber \\
w &\equiv & [Y_L(\hat k) \otimes Y_{L'}(\hat q)]^2 \cdot [\hat k \otimes \hat q]^2 \ .
\label{a2}
\end{eqnarray}
In order to arrive at Eq.(\ref{a1}), we have also made use of the results
\begin{eqnarray}
[\hat q \otimes \hat q]^2 \cdot [\hat q \otimes \hat q]^2  & = & \frac{2}{3} \nonumber \\
\small[\hat k \otimes \hat k]^2 \cdot [\hat q \otimes \hat q]^2  & = & \frac{1}{3} (3\cos^2\theta-1) \nonumber \\
\small[\hat k \otimes \hat q]^2 \cdot [\hat q \otimes \hat q]^2  & = & \frac{2}{3} \cos\theta \nonumber \\
\small[\hat k \otimes \hat q]^2 \cdot [\hat k \otimes \hat q]^2  & = & \frac{1}{6} (3+\cos^2\theta) \ .
\label{a3}
\end{eqnarray}
Eq.(\ref{a1}) can be readily inverted to yield
\begin{eqnarray}
a_{L'L}  & = & \frac{1}{\sin^4\theta} \left[2 u + (1+\cos^2\theta) v - 4\cos\theta w \right]
 \nonumber \\
b_{L'L}  & = & \frac{1}{\sin^4\theta} \left[(1+\cos^2\theta) u + 2 v - 4\cos\theta w \right]
 \nonumber \\
c_{L'L}  & = & \frac{2}{\sin^4\theta} \left[-2\cos\theta (u + v)  + (3\cos^2\theta + 1) w \right] \ .
\label{a4}
\end{eqnarray}

Choosing the quantization axis $\hat z$ along $\hat k$, the quantities $u$, $v$, and $w$ 
defined in Eq.(\ref{a2}) can be expressed without loss of generality as
\begin{eqnarray}
u & = & \frac{1}{4\pi} \sqrt{\frac{2}{3}} [LL'] (L0L'0|20) P_L(\hat k \cdot \hat q) \nonumber \\
v & = & \frac{1}{4\pi} \sqrt{\frac{2}{3}} [LL'] (L0L'0|20) P_{L'}(\hat k \cdot \hat q) \nonumber \\
w & = & \frac{[LL']}{\sqrt{4\pi}} \sum_l \frac{1}{[l]} (L'010|l0) 
\sum_{M,m_l} (L0L'M|2M)(101M|2M) (L'M1M|lm_l) Y_{lm_l}(\theta, 0) \ .
\label{a5}
\end{eqnarray}

Similarly to what has been done above, taking the scalar product of the last equality in
Eq.(\ref{aux11n}) with $[\hat k \otimes \hat n_2]^2$ and $[\hat q \otimes \hat n_2]^2$, 
respectively, we have
\begin{eqnarray}
2 r & = & a'_{L'L} + \cos\theta b'_{L'L} \nonumber \\
2 t & = & \cos\theta a'_{L'L} + b'_{L'L}  \ ,
\label{a6}
\end{eqnarray}
where
\begin{eqnarray}
r &\equiv & [Y_L(\hat k) \otimes Y_{L'}(\hat q)]^2 \cdot [\hat k \otimes \hat n_2]^2 \nonumber \\
t &\equiv & [Y_L(\hat k) \otimes Y_{L'}(\hat q)]^2 \cdot [\hat q \otimes \hat n_2]^2 \ .
\label{a7}
\end{eqnarray}
In order to arrive at Eq.(\ref{a6}), we have also made use of the results
\begin{eqnarray}
[\hat k \otimes \hat n_2]^2 \cdot [\hat k \otimes \hat n_2]^2  & = & \frac{1}{2} \nonumber \\
\small[\hat k \otimes \hat n_2]^2 \cdot [\hat q \otimes \hat n_2]^2  & = & \frac{1}{2} \cos\theta \ .
\label{a8}
\end{eqnarray}
Eq.(\ref{a6}) can be readily inverted to yield
\begin{eqnarray}
a'_{L'L}  & = & \frac{2}{\sin^2\theta} \left(r - t \cos\theta \right) \ , \nonumber \\
b'_{L'L}  & = & \frac{2}{\sin^2\theta} \left(t - r \cos\theta \right) \ .
\label{a9}
\end{eqnarray}

Choosing the quantization axis $\hat z$ along $\hat k$, the quantities $r$ and $t$
defined in Eq.(\ref{a7}) can be expressed as
\begin{eqnarray}
r & = & - i \frac{1}{4\pi} \frac{[LL']}{\sqrt{L'(L'+1)}} (L0L'1|21) P^1_{L'}(\hat k \cdot \hat q) \nonumber \\
t & = & - i \frac{[LL']}{\sqrt{2\pi}} \sum_l \frac{1}{[l]} (L'010|l0)
\sum_{M,m's} (-)^M (L0L'M|2M)(L'M1m|lm_l)  \nonumber \\
&\times & (1m11|2-M) Y_{lm_l}(\theta, 0) \ .
\label{a10}
\end{eqnarray}
\section{Appendix B}
In this appendix we will give the explicit expression for the quantities 
$A_{L'}$, $B_{L'}$ and $C_{L'}$ in Eq.(\ref{SSTRUC1_P_thr}) and $A'_{L'}$, 
$B'_{L'}$ and $C'_{L'}$ in Eq.(\ref{SSTRUC1_N_thr}).

Using Eqs.(\ref{SSTRUC12a},\ref{SSTRUC12_1Pa}) for those states specified in 
Eq.(\ref{SPpw_p1}), we find
\begin{eqnarray}
A_0 & = & \frac{1}{4\pi\sqrt{3}} \left[ M^{\frac{1}{2}\frac{1}{2}}_{00} 
          - \sqrt{2} M^{\frac{1}{2}\frac{3}{2}}_{02}\right] \ , \nonumber \\
\left(\frac{q}{\Lambda}\right) A_1 & = & \frac{1}{4\pi}  
\left(\frac{\sqrt{2}-1}{\sqrt{2}}\right) \sqrt{\frac{2}{3}} 
\left[M^{\frac{1}{2}\frac{1}{2}}_{11} - M^{\frac{3}{2}\frac{1}{2}}_{11}
+ \frac{1}{2} M^{\frac{1}{2}\frac{3}{2}}_{11} + \sqrt{\frac{5}{2}} M^{\frac{3}{2}\frac{3}{2}}_{11} 
\right] \nonumber \\
& + & \frac{1}{4\pi} \sqrt{\frac{2}{3}} 
\left[M^{\frac{1}{2}\frac{3}{2}}_{11} + M^{\frac{3}{2}\frac{3}{2}}_{11} \right] 
- \frac{1}{4\pi} \sqrt{\frac{3}{5}} M^{\frac{3}{2}\frac{3}{2}}_{13} 
\ , \nonumber \\ 
\left(\frac{q}{\Lambda}\right) B_1 & = & \frac{1}{4\pi}  
\left(\frac{\sqrt{2}+1}{\sqrt{2}}\right) \sqrt{\frac{2}{3}} 
\left[M^{\frac{1}{2}\frac{1}{2}}_{11} - M^{\frac{3}{2}\frac{1}{2}}_{11}
+ \frac{1}{2} M^{\frac{1}{2}\frac{3}{2}}_{11} + \sqrt{\frac{5}{2}} M^{\frac{3}{2}\frac{3}{2}}_{11} 
\right] \ , \nonumber \\
\left(\frac{q}{\Lambda}\right) C_1 & = & - \frac{1}{4\pi}  
\sqrt{\frac{3}{8}} 
\left[M^{\frac{1}{2}\frac{3}{2}}_{11} + \sqrt{2} M^{\frac{3}{2}\frac{3}{2}}_{11}
- 2\sqrt{3} M^{\frac{3}{2}\frac{3}{2}}_{13}\right] \ . 
\label{b1}
\end{eqnarray}

Similarly, using Eqs.(\ref{SSTRUC12na},\ref{SSTRUC_1Pan}) for those states specified in 
Eq.(\ref{SPpw_n1}), we find
\begin{eqnarray}
\left(\frac{q}{\Lambda}\right) A'_1 & = & - \frac{1}{4\pi}  
\left(\frac{\sqrt{2}+1}{\sqrt{2}}\right) \sqrt{\frac{1}{6}} 
\left[M^{\frac{1}{2}\frac{1}{2}}_{10} + 2 M^{\frac{1}{2}\frac{3}{2}}_{12}
- 4\sqrt{\frac{2}{3}} M^{\frac{3}{2}\frac{3}{2}}_{10} - \sqrt{2} M^{\frac{3}{2}\frac{3}{2}}_{12} 
+ 4\sqrt{\frac{1}{3}} M^{\frac{3}{2}\frac{3}{2}}_{12} \right] \ , \nonumber \\
B'_0 & = & \frac{1}{4\pi}\left(\frac{\sqrt{2}+1}{\sqrt{2}}\right) \sqrt{\frac{1}{3}}
\left[M^{\frac{1}{2}\frac{1}{2}}_{01} - M^{\frac{1}{2}\frac{3}{2}}_{01}\right] \ , \nonumber \\
\left(\frac{q}{\Lambda}\right) B'_1 & = & - \frac{1}{4\pi}
\left(\frac{\sqrt{2}+1}{\sqrt{2}}\right) \sqrt{\frac{1}{3}}
\left[M^{\frac{1}{2}\frac{1}{2}}_{10} + M^{\frac{1}{2}\frac{3}{2}}_{12}
+ \sqrt{2} M^{\frac{3}{2}\frac{3}{2}}_{10} - 2 M^{\frac{3}{2}\frac{1}{2}}_{12} 
\right] \ , \nonumber \\
\left(\frac{q}{\Lambda}\right) C'_1 & = & - \frac{1}{4\pi} \sqrt{\frac{3}{8}} 
\left[M^{\frac{1}{2}\frac{3}{2}}_{12} + \sqrt{\frac{1}{2}} M^{\frac{3}{2}\frac{3}{2}}_{12}
\right] \ . 
\label{b2}
\end{eqnarray}
\section{Appendix C}
The quantities $A^S_{L'}$, $B^S_{L'}$, etc, in Eq.(\ref{SSTRUC1_pam_pw}) are given by
\begin{eqnarray}
A^0_0 & = & \frac{1}{4\pi} M^{000}_{00} \ , \nonumber \\
\left(\frac{p'}{\Lambda}\right) A^1_1 & = & 
\frac{1}{4\pi} \frac{1}{3} \left[M^{110}_{11} + 3 M^{111}_{11} + 5 M^{112}_{11} \right]
 \ , \nonumber \\
\left(\frac{p'}{\Lambda}\right) B^1_1 & = & -
\frac{1}{4\pi} \frac{1}{4} \left[M^{110}_{11} + \frac{3}{2} M^{111}_{11} 
- \frac{5}{2} M^{112}_{11} \right] \ , \nonumber \\
\left(\frac{p'}{\Lambda}\right) C^1_1 & = &
\frac{1}{4\pi} \frac{3}{2\sqrt{2}} M^{011}_{11} \ , \nonumber \\
\left(\frac{p'}{\Lambda}\right) D^1_1 & = & -
\frac{1}{4\pi} \sqrt{\frac{3}{2}} \frac{5}{2} M^{112}_{13} \ , \nonumber \\
\left(\frac{p'}{\Lambda}\right) E^1_1 & = & 
\frac{1}{4\pi} \left[ \sqrt{\frac{3}{2}} M^{112}_{13} 
- \frac{1}{2} M^{110}_{11} + \frac{3}{4} M^{111}_{11} - \frac{1}{4} M^{112}_{11}
\right] \ .
\label{SSTRUC1_pam1}
\end{eqnarray}

For convenience we define
\begin{eqnarray}
\left(\frac{p'}{\Lambda}\right) U^1_1 & \equiv & 
\frac{\sqrt{3}}{4\pi} \left[\sqrt{\frac{3}{2}} \left(M^{112}_{11} - M^{111}_{11}\right) + M^{112}_{11} \right]
 \ , \nonumber \\
\left(\frac{p'}{\Lambda}\right) W^1_1 & \equiv & 
\frac{\sqrt{3}}{4\pi} \left[\sqrt{\frac{3}{2}} \left(M^{112}_{11} + M^{111}_{11}\right) + M^{112}_{11} \right]
 \ .
\label{SSTRUC1_pam11}
\end{eqnarray}

Similarly, the quantities $A'^{S'S}_{L'}$, $B'^{S'S}_{L'}$, etc, in Eq.(\ref{SSTRUC1_nam_pw}) 
are given by
\begin{eqnarray}
A'^{11}_0 & = & 
\frac{1}{4\pi} \frac{1}{2}\sqrt{\frac{3}{2}} M^{111}_{01} \ , \nonumber \\
\left(\frac{p'}{\Lambda}\right) B'^{10}_1 & = & 
\frac{1}{4\pi} \frac{1}{2} \left[M^{100}_{10} - \sqrt{\frac{5}{2}} M^{102}_{12}\right]  
\ , \nonumber \\
\left(\frac{p'}{\Lambda}\right) C'^{10}_1 & = & 
\frac{1}{4\pi} \frac{1}{2} \left[M^{100}_{10} + \sqrt{10} M^{102}_{12}\right] \ , \nonumber \\
C'^{01}_0 & = & 
\frac{1}{4\pi} \frac{1}{2} M^{010}_{01} \ .
\label{SSTRUC1_nam1}
\end{eqnarray}
\section{Appendix D}

In order to calculate the observables directly from the transition operators in 
Eqs.(\ref{SSTRUC1p},\ref{SSTRUC1n},\ref{SSTRUC2_MNN}), it is convenient to express 
them in the form
\begin{equation}
\hat M = \sum_\lambda \sum_{n,n'=0}^3 
M^\lambda_{nn'} \sigma_n(1) \sigma_{n'}(2) \ ,
\label{ampl_o}
\end{equation}
where $\sigma_0(i) \equiv 1$, $\sigma_1(i) \equiv \sigma_x(i)$, etc., for $i$-th nucleon. 

The unpolarized cross section is then given by
\begin{eqnarray}
\frac{d\sigma}{d\Omega} & \equiv & \frac{1}{4} Tr[\hat M \hat M^\dagger] \nonumber \\
& = & \sum_{\lambda,\lambda'} \sum_{n,n'=0}^3 M^\lambda_{nn'}(M^{\lambda'}_{nn'})^* \ .
\label{xsc}
\end{eqnarray}

For the analyzing power we have 
\begin{eqnarray}
\frac{d\sigma}{d\Omega} A_i & \equiv & \frac{1}{4} Tr[\hat M \sigma_i(1) \hat M^\dagger] \nonumber \\
& = & \sum_{\lambda,\lambda'} \sum_{n'=0}^3 \left\{ 2\Re[M^\lambda_{in'}(M^{\lambda'}_{0n'})^*]
+ i \sum_{k,n=1}^3 \varepsilon_{kni} M^\lambda_{nn'}(M^{\lambda'}_{kn'})^*  
\right\} \ ,
\label{Ai}
\end{eqnarray}
where $\varepsilon_{kni}$ denotes the Levi-Civita antisymmetric tensor. We note that, 
from parity conservation, $A_x=A_z=0$ for the two-body reaction $NN\to B'B$; this result 
also holds for the three-body reaction $NN\to MB'N$ in the coplanar geometry.

The spin correlation coefficient $A_{ij}$ is given by
\begin{eqnarray}
\frac{d\sigma}{d\Omega} A_{ij} & \equiv & \frac{1}{4} Tr[\hat M \sigma_i(1) \sigma_j(2) \hat M^\dagger] \nonumber \\
& = & \sum_{\lambda,\lambda'} \Big\{ 
2\Re[ M^\lambda_{00}(M^{\lambda'}_{ij})^* + M^\lambda_{i0}(M^{\lambda'}_{0j})^* ]
 +  \sum_{k,n=1}^3 2\Im[ M^\lambda_{0k}(M^{\lambda'}_{in})^* \varepsilon_{knj} 
    + M^\lambda_{k0}(M^{\lambda'}_{nj})^* \varepsilon_{kni}] \nonumber \\
& - & \sum_{k,k',n,n'=1}^3 M^\lambda_{kk'}(M^{\lambda'}_{nn'})^* \varepsilon_{nki}\varepsilon_{n'k'j} 
\Big\} \ .
\label{Cij}
\end{eqnarray}

Of course, any observable can also be expressed in terms of the spin-matrix elements
given by Eqs.(\ref{PWE},\ref{PWE_MNN}). We have, e.g., 
\begin{equation}
\frac{d\sigma}{d\Omega} =  
\frac{1}{4} \sum_{S,S',M_S,M_S'}|<S'M_S'|\hat M|SM_S>|^2 \ ,
\label{xsc_SME}
\end{equation}
\begin{eqnarray}
\frac{d\sigma}{d\Omega} A_x & = & \frac{\sqrt{2}}{4} \sum_{S,S',M_S'} 
\Re[ <S'M_S'|\hat M|S0><S'M_S'|\hat M|1-1>^* \nonumber \\
&+& (-)^{1+S}<S'M_S'|\hat M|11><S'M_S'|\hat M|S0>^*] \ , \nonumber \\
\frac{d\sigma}{d\Omega} A_y & = & \frac{\sqrt{2}}{4} \sum_{S,S',M_S'} 
\Im[ <S'M_S'|\hat M|S0><S'M_S'|\hat M|1-1>^* \nonumber \\
&+& (-)^{1+S}<S'M_S'|\hat M|11><S'M_S'|\hat M|S0>^*] \ , \nonumber \\
\frac{d\sigma}{d\Omega} A_z & = & \frac{1}{4} \sum_{S',M_S'}
\left[ |<S'M_S'|\hat M|11>|^2 - |<S'M_S'|\hat M|1-1>|^2 \right] \ ,
\label{Ai_SME}
\end{eqnarray}
\begin{eqnarray}
\frac{d\sigma}{d\Omega} A_{xx} & = & \frac{1}{4} \sum_{S,S',M_S,M_S'}
(-)^{1+S-2M_S}<S'M_S'|\hat M|SM_S><S'M_S'|\hat M|S-M_S>^*  \ , \nonumber \\
\frac{d\sigma}{d\Omega} A_{yy} & = & \frac{1}{4} \sum_{S,S',M_S,M_S'}
(-)^{1+S-M_S}<S'M_S'|\hat M|SM_S><S'M_S'|\hat M|S-M_S>^*  \ , \nonumber \\
\frac{d\sigma}{d\Omega} A_{zz} & = & \frac{1}{4} \sum_{S,S',M_S,M_S'}
(-)^{1+M_S}|<S'M_S'|\hat M|SM_S>|^2 \ .
\label{Cij_SME}
\end{eqnarray}

As has been pointed out in Ref.\cite{Bilenky}, a given initial state spin 
contribution to the cross section can be isolated by measuring some spin 
observables. In particular, using the spin-projection operator $P_S$ as 
given in section III in terms of the Pauli spin matrices, it is immediate 
that
\begin{eqnarray}
\frac{d(^1\sigma)}{d\Omega} &\equiv &  \frac{1}{4}Tr[(\hat M P_{S=0})(\hat M P_{S=0})^\dagger] 
= \frac{1}{4}Tr[\hat M P_{S=0}\hat M^\dagger] \nonumber \\
&=& \frac{1}{4}\frac{d\sigma}{d\Omega}(1 - A_{xx} - A_{yy} - A_{zz}) \ , \nonumber \\
\frac{d(^3\sigma)}{d\Omega} & \equiv &  \frac{1}{4}Tr[(\hat M P_{S=1})(\hat M P_{S=1})^\dagger] 
= \frac{1}{4}Tr[\hat M P_{S=1}\hat M^\dagger] \nonumber \\
&=& \frac{1}{4}\frac{d\sigma}{d\Omega} 
(3 + A_{xx} + A_{yy} + A_{zz}) \ , 
\label{S_xsc4}
\end{eqnarray}
where $\, ^{2S+1}\sigma$ denotes the (initial state) spin-singlet or -triplet 
cross section as $S=0$ or $1$, respectively.

Also, Eq.{\ref{Cij_SME} can be inverted to solve for the spin cross 
sections, 
$d(\,^{2S+1}\sigma_{M_S})/d(\Omega) \equiv \sum_{S',M_S'} |<S'M_S'|\hat M|SM_S>|^2 / 4$.
We obtain \cite{Meyer1}
\begin{eqnarray}
\frac{d(^1\sigma_0)}{d\Omega} & = & \frac{1}{4}\frac{d\sigma}{d\Omega} 
(1 - A_{xx} - A_{yy} - A_{zz}) \ , \nonumber \\
\frac{d(^3\sigma_0)}{d\Omega} & = & \frac{1}{4}\frac{d\sigma}{d\Omega} 
(1 + A_{xx} + A_{yy} - A_{zz}) \ , \nonumber \\
\frac{d(^3\sigma_1)}{d\Omega} + \frac{d(^3\sigma_{-1})}{d\Omega} 
                             & = & \frac{1}{2}\frac{d\sigma}{d\Omega} 
(1 + A_{zz}) \ ,
\label{S_xsc1}
\end{eqnarray}
where we have made used of the relation
$\sigma = \, ^1\sigma_0 + \, ^3\sigma_0 + \, ^3\sigma_1 + \, ^3\sigma_{-1}$.
Note that Eq.(\ref{S_xsc1}) is consistent with Eq.(\ref{S_xsc4}) as 
$\, ^1\sigma = \, ^1\sigma_0$ and
$\, ^3\sigma = \, ^3\sigma_0 + \, ^3\sigma_1 + \, ^3\sigma_{-1}$.

From Eq.(\ref{S_xsc1}), it is immediate that the spin-triplet cross section defined
as 
\begin{equation}
\frac{d(^3\sigma_\Sigma)}{d\Omega} \equiv \frac{d(^3\sigma_0)}{d\Omega}
+ \frac{1}{2}\left(\frac{d(^3\sigma_1)}{d\Omega}+ \frac{d(^3\sigma_{-1})}{d\Omega}\right)
\label{xsc30}
\end{equation}
is given by Eq.(\ref{xsc3}). Note that $d(^3\sigma_1)/d\Omega = d(^3\sigma_{-1})/d\Omega$ for 
$NN\to B'B$ by symmetry. This holds also for $NN\to MB'N$ in the coplanar geometry
or for the cross section integrated over the emission angle of the one of the three 
particles in the final state.


%

\begin{thebibliography}{99}
\bibitem{notation}
Here we adopt the same notation for the partial-wave states as used in Ref.~\cite{Meyer}, 
i.e., $^{2S+1}L_J \rightarrow \ ^{2S'+1}L'_{J'} l$.
\end{thebibliography}
\end{document}